\documentclass{ieeeaccess}
\usepackage{amsmath,amssymb,amsfonts}
\usepackage{algorithmic}
\usepackage{graphicx}
\usepackage{textcomp}

\usepackage{algorithm}
\usepackage{algorithmic}
\usepackage[caption=false]{subfig}
\usepackage{bm}
\usepackage{multirow}
\usepackage{comment}
\usepackage{url}
\usepackage{here}

\newcommand{\logdet}{\mathrm{logdet}}

\newcommand{\prox}{\mathrm{prox}}
\newcommand{\soft}{\mathrm{soft}}
\newcommand{\diag}{\text{diag}}

\newtheorem{theorem}{Theorem} 
\newtheorem{proposition}[theorem]{Proposition}

\usepackage[
backend=biber,
style=ieee,
sorting=none,
isbn=false,
url=false, 
doi=false,
eprint=false
]{biblatex}

\def\BibTeX{{\rm B\kern-.05em{\sc i\kern-.025em b}\kern-.08em
    T\kern-.1667em\lower.7ex\hbox{E}\kern-.125emX}}
\begin{document}

\history{}
\doi{}


\title{Learning Sparse Graph with Minimax Concave Penalty under Gaussian Markov Random Fields}
\author{\uppercase{TATSUYA KOYAKUMARU}\authorrefmark{1},
\uppercase{Masahiro Yukawa}\authorrefmark{1},
\uppercase{Eduardo Pavez}\authorrefmark{2},
and \uppercase{Antonio Ortega}.\authorrefmark{2}}
\address[1]{Department of Electronics and Electrical Engineering, Keio
  University, Kanagawa 223-8522, Japan}
\address[2]{Department of Electrical and Computer Engineering, University of Southern California, Los Angeles, California, CA 90089, USA}

\tfootnote{This work was supported by the Grants-in-Aid for Scientific Research (KAKENHI) under Grant JP18H01446.}

\markboth
{T.Koyakumaru \headeretal: Learning Sparse Graph with Minimax Concave Penalty under Gaussian Markov Random Fields}
{T.Koyakumaru \headeretal: Learning Sparse Graph with Minimax Concave Penalty under Gaussian Markov Random Fields}

\corresp{Corresponding author: Masahiro Yukawa (e-mail: yukawa@elec.keio.ac.jp).}

\begin{abstract}
This paper presents a convex-analytic framework to learn sparse graphs from data.
    While our problem formulation is inspired by an extension of the graphical lasso using the so-called combinatorial graph Laplacian framework, a key difference is the use of 
    a nonconvex alternative to the $\ell_1$ norm to attain graphs with better interpretability.
    Specifically, we use the weakly-convex minimax concave penalty (the difference between the $\ell_1$ norm and the Huber function) which is known to yield sparse solutions with lower estimation bias than $\ell_1$ for regression problems.
    In our framework, the graph Laplacian is replaced in the optimization by a linear transform of the vector corresponding to its upper triangular part.
    Via a reformulation relying on Moreau's decomposition, we show that  overall convexity is guaranteed by introducing a quadratic function to our cost function.
    The problem can be solved efficiently by
    the primal-dual splitting method, of which the admissible conditions for provable convergence are presented.
    Numerical examples show that the proposed method significantly outperforms the existing graph learning methods with reasonable CPU time.
\end{abstract}

\begin{keywords}
Graph signal processing, graph learning, graphical lasso, minimax concave penalty, primal-dual splitting method, proximity operator
\end{keywords}

\titlepgskip=-15pt

\maketitle

\section{Introduction}
\label{sec:intro}
How can we learn sparse graphs with enhanced interpretability under the Gaussian Markov random field (GMRF) \cite{GMRF}?
This is the central question addressed in this article.
A graph, containing a set of vertices and edges, is a mathematical tool to represent the dependencies among components (such as nodes of a network or pixels of an image), through the selection of pairwise relations (edge weights) between each pair of objects (vertices).
In particular, the strength of the relation can be expressed in terms of (nonnegative) graph weights.
In the present context, graph ``sparseness'' is an important property because it tends to provide better interpretability, i.e., relative to all possible connections between nodes, only a few edges are non-zero and provide information about the major relationships between objects.  

The problem of learning graphs from data 
has been studied widely in a variety of fields including signal processing, machine learning, and statistics 
\cite{friedman2008sparse, mazumder2012,Meinshausen_06,JMLR:v9:banerjee08a,GSP_GL_2,wang2012, Sun12, Lake,ortega16,palomar20,lam2009}.
Graph learning has been considered in multiple applications such as design of 
functional brain network architectures \cite{VECCHIO2017206}, 
molecular biology \cite{Mason2007}, and network anomaly detection \cite{Bhuyan16}.
We also refer the reader to \cite{connect, Dong_servey} for comprehensive reviews of graph learning. 
The graphical model approach \cite{friedman2008sparse, mazumder2012,Meinshausen_06,JMLR:v9:banerjee08a,GSP_GL_2}
represents dependencies with the data in graph form and 
has gained significant popularity owing to two main reasons.
First, the graphical model is built upon a solid statistical foundation, so that the edge weights have a physical meaning  under certain assumptions. For instance, if the observed data are derived from a GMRF model, the weights are based on partial correlation coefficients \cite{book_Probabilistic_graphical_model}.
Second, it provides excellent versatility as it assumes no specific structure on the graph.
A particular example of graph learning algorithm is
the graphical lasso \cite{friedman2008sparse,Bhuyan14, wang2012, Sun12, Lake}, which
employs  $\ell_1$ regularization on the edge weights
to obtain the  sparse inverse covariance matrix of a GMRF model \cite{GMRF}. 
This approach has been extended and modified in \cite{ortega16} to learn several types of  Laplacian matrices, including a formulation where the inverse covariance matrix has a combinatorial graph Laplacian (CGL) structure \cite{ortega16}.
As noted earlier, research on sparse graph learning is motivated by the fact that sparsity enhances the interpretability of the learned graphs  \cite{Buhlmann_book,ortega16}.
All those sparsity-seeking methods exploit convex penalties (such as the $\ell_1$ norm) mainly
due  to their mathematical tractability.

To clarify the motivation of the present study, let us turn our attention to sparse linear regression.
A plethora of nonconvex alternatives to the $\ell_1$ regularization have been proposed 
to reduce the estimation bias while maintaining the benefit of variance reduction
\cite{zhang2010, surveyOfNonconvexPenalty, Mazumder_11, lam2009}.
Among them, we focus on the minimax concave (MC) penalty \cite{zhang2010, selesnick} 
because: 
\begin{enumerate}
    \item[(i)] it saturates (i.e., it returns a constant value when the variable being estimated exceeds a given threshold) thereby reducing the estimation biases significantly;  
        \item[(ii)] it has been shown to bridge the gap between the $\ell_0$ and $\ell_1$ norms in a parametric way \cite{Abe_2020};  
            \item[(iii)] it is a weakly convex function \cite{Nurminskii_73_weakly_convex}; more specifically, it is given by subtracting from the $\ell_1$ norm its Moreau envelope.
\end{enumerate}
Property (iii) is of particular importance
from an optimization viewpoint because the overall convexity of the cost function 
is ensured when the MC penalty is used with  strongly convex loss functions, and also because
the decomposed form in terms of a convex function and its Moreau envelope is compatible with the efficient operator splitting methods.
The MC penalty has been used in various sparse estimation problems, e.g., feature selection with a sparse support vector machine (SVM) \cite{ex_MCP_1_Feature_Selection} and gear fault diagnosis from noisy vibration signals \cite{ex_MCP_2_Bearing_Fault_Diagnosis}.
To the best of authors' knowledge, the $\ell_p$ quasi-norm for $p\in(0,1)$ is the only function, excluding the MC penalty, that is known to possess property (ii) above, but it lacks properties (i) and (iii).
On the other hand, the smoothly clipped absolute deviation (SCAD) penalty \cite{lam2009} is similar to the MC penalty and it could be an alternative choice, since 
it actually possesses property (i) as well as the weak-convexity part of property (iii) although further investigations would be needed to determine whether property (ii) and the other part of property (iii) also hold for the SCAD penalty.
While nonconvex alternatives to the $\ell_1$ penalty have been successful in the context of sparse linear regression, their study for graph learning has been limited,
and most of the existing graph learning methods use the $\ell_1$ or $\ell_2$ regularization. 
A few exceptions include approaches using the log function \cite{palomar20} or
the $\ell_0$ norm \cite{Shen12}.
A use of the SCAD penalty was also mentioned in \cite{lam2009, palomar_MC}.
The recent works \cite{palomar_MC, zhang2020learning} have observed that an increase of the regularization parameter of the $\ell_1$ penalty in the CGL estimation framework ultimately does not lead to a sparse solution and instead produces  a dense solution associated with a fully connected graph.
Based on this observation, in \cite{palomar_MC, zhang2020learning},
it has been shown that the use of the MC penalty (as well as other nonconvex penalties)
yields better performance.
However, these approaches are based on a nonconvex formulation and thus there is no guarantee that the generated sequence of graphs converges to a global optimum.
This motivates us to devise another formulation which benefits from the weak convexity of the MC penalty to guarantee overall convexity of the entire cost so that the generated graphs converge to provably global optimum.

The goal of this article is to present a novel graph learning framework based on
a convex formulation involving the nonconvex (but weakly convex) MC penalty
to produce sparse graphs, and specifically sparse CGL matrices.
Since CGLs are symmetric matrices, we remove this redundancy   by 
representing a CGL matrix using a linear transform of the  vector of graph weights corresponding to the upper-triangular part.\footnote{Although the one-to-one linear operator for representing the CGL was used in the structured graph learning via Laplacian spectral constraints (SGL) method \cite{palomar20}, the proposed method is more efficient (as shown in Section \ref{numerical_example}) due to the proposed
reformulation, which allows to use the primal-dual splitting method \cite{primal-dual} (as shown in Section \ref{proposed_algorithms}).}
Here, the upper-triangular part represents the undirected relations among nodes and completely characterizes the CGL matrix, so that  our estimate is automatically guaranteed to have a Laplacian structure without the need to impose any constraints. 
This is in sharp contrast to the existing CGL approaches, which typically require both a  positive semi-definite constraint and a linear constraint.
Our formulation involves the nonconvex MC penalty, instead of the $\ell_1$ norm,
while essentially keeping the same terms
(the ``nonsmooth'' log-determinant term and the linear term) as the graphical lasso formulation
but with the linear operator mentioned above.
Note here that the negative log-determinant function is differentiable but with non-Lipschitz-continuous gradient.
Due to the nonconvexity and the nonsmoothness, the problem cannot be solved directly using existing optimization methods.
To circumvent the difficulty, we invoke the classical Moreau's decomposition and show
that the Tikhonov regularization convexifies the overall cost function,
reformulating the problem into a canonical form of the primal-dual splitting method \cite{primal-dual}.
We present the admissible conditions under which the convergence to the global optimal point is guaranteed by the primal-dual splitting method.
Numerical examples show that the proposed method outperforms the conventional CGL method (its $\ell_1$-based counterpart) for three types of graph.
Compared to the state-of-the-art method,
the structured graph learning via Laplacian spectral constraints (SGL) \cite{palomar20}, 
the proposed method achieves comparable or better performance, depending on the type of graph, with up to 40 times shorter CPU time.
In addition, experiments with real data show that the method produces a sparser graph 
than other existing methods.

New features of the present work relative to our preliminary work \cite{koyakumaru_21}
include
detailed proofs of the mathematical results and refined experimental results as well as
additional simulation results using real data.

\section{PRELIMINARIES}
We present notation, and then show some mathematical tools used in this work. We finally present the primal-dual splitting method which is used to solve the proposed optimization problem to be  presented in Section \ref{proposed_algorithms}.

\subsection{Notation}
The sets of real numbers and nonnegative real numbers are denoted by $\mathbb{R}$ and $\mathbb{R}_+$, respectively.
The transpose of vector/matrix is denoted by $(\cdot)^T$. 
Given a vector $\bm{x}:=[x_1,x_2,\cdots,x_n]^T$ $\in \mathbb{R}^n$,
define the $\ell_1$ and the $\ell_2$ norms by
$\|\bm{x}\|_1$ := $\sum_{i=1}^n$ $| x_i|$ and
$\|\bm{x}\|_2$ := $ \left(\sum_{i=1}^n x_i^2\right)^\frac{1}{2}$, respectively.
Similarly, given a matrix $\bm{X} \in \mathbb{R}^{n\times n}$ with its $(i,j)$ component denoted by $x_{i,j}$,  define the $\ell_1$ and the Frobenius norms by
$\|\bm{X}\|_1:=\sum_{i,j=1}^{n}| x_{i, j}|$ and
$\|\bm{X}\|_{\rm F} := \left(\sum_{i,j=1}^{n} x_{i, j}^2\right)^\frac{1}{2}$, respectively.
Given a pair of matrices $\bm{A}$ and $\bm{B}$,
define the inner product $\langle \bm{A}, \bm{B} \rangle := \sum_{i, j=1}^{n}a_{i, j}b_{i, j}$. Let $\bm{I}$ and $\bm{1}$ denote the identity matrix and  the vector of ones, respectively, 
and let $\text{diag}(\bm{x})$ represent the diagonal matrix consisting of the components of a vector $\bm{x}$.

We consider undirected weighted graphs with nonnegative edge weights. 
The graph $\mathcal{G}=(\mathcal{V}, \mathcal{E}, \bm{W})$ is composed of a set of nodes $\mathcal{V}$, edges $\mathcal{E}$, and a symmetric weight matrix $\bm{W}\in\mathbb{R}^{n\times n}$ with $w_{i,j} > 0$ if $(i,j) \in \mathcal{E}$, and $w_{i,j} = 0$ if $(i,j) \not\in \mathcal{E}$, where $n=|\mathcal{V}|$ is the number of nodes.
Here, $(i,i)\not\in \mathcal{E}$ for any $i$ by convention.
CGL is defined by $\bm{\Theta}=\bm{D}-\bm{W}\in\mathbb{R}^{n\times n}$, where
$\bm{D}:=\text{diag}(\bm{W1})$ is the degree matrix.
CGL has zero row-sums with its minimum eigenvalue also zero which is simple when the graph is connected. 

\subsection{Mathematical tools}
The conjugate of a function $f(\bm{w})$ is denoted by $f^*(\bm{y})=\sup_{\bm{w} \in \mathbb{R}^N}\langle \bm{w}, \bm{y} \rangle - f(\bm{w})$, $\bm{y}\in\mathbb{R}^N$.
The set of proper lower semicontinuous convex functions from $\mathbb{R}^N$ to ($-\infty, +\infty$] is denoted by $\Gamma_0(\mathbb{R}^N)$.\footnote{
A function $f$  is {\it proper} if $\operatorname{dom} f:=\{\boldsymbol{w} \in 
\left.\mathbb{R}^{N} \mid f(\boldsymbol{w})<+\infty\right\} \neq \emptyset$, and {\it lower semicontinuous} at $\boldsymbol{w}$  if $f(\boldsymbol{w}) \leq
\lim \inf _{\boldsymbol{y} \rightarrow \boldsymbol{w}} \!f(\boldsymbol{y})$.
}
 The proximity operator of $f\in\Gamma_0(\mathbb{R}^N)$ of index $\gamma > 0$ is defined as follows \cite{combettes_book}: 
\begin{equation}
    \operatorname{prox}_{\gamma f} (\bm{w}):= \underset{\bm{y} \in \mathbb{R}^{N}}{\operatorname{argmin}}\left(f(\bm{y})+\frac{1}{2\gamma}\|\bm{w}-\bm{y}\|^2_2 \right).
\end{equation}
Uniqueness and existence of the minimizer is guaranteed by the strong convexity and coercivity of $f + \frac{1}{2\gamma} \|\bm{w}-\cdot\|^2_2$. 
The indicator function with respect to a given set $\mathcal{S}$ is denoted by 
\begin{equation}
    \iota_\mathcal{S} (\bm{w}):=\begin{cases}0,&\mbox{if }\bm{w}\in \mathcal{S},\\ +\infty, & \mbox{otherwise}.\end{cases}
\end{equation}
It is clear by definition that
$\prox_{\iota_C}(\bm{w}) = P_{C}(\bm{w}) := 
\underset{\bm{y} \in C} {\operatorname{argmin}} 
\|\bm{w} - \bm{y}\|_2$.
The Moreau envelope of a function $f \in \Gamma_0(\mathbb{R}^N)$ of index $\gamma > 0 $ is defined as follows \cite[Definition 12.20]{combettes_book}: 
\begin{eqnarray}
 ^\gamma \!f(\bm{w})&:=&\min _{\boldsymbol{y} \in \mathbb{R}^{N}}\left(f(\boldsymbol{y})+\frac{1}{2\gamma}\|\boldsymbol{w}-\boldsymbol{y}\|_2^2\right).
	\label{infimal_convolution}
\end{eqnarray}
	
\noindent Using the Moreau envelope $^\gamma \|\cdot \|_1$ of $\|\cdot\|_1$, which is the widely known Huber function, the MC penalty \cite{selesnick} is defined  as
\begin{eqnarray} 
\phi_{\mathrm{MC}}(\bm{w})
&=&\|\bm{w}\|_1-^\gamma \!\|\cdot\|_1(\bm{w}) . \label{MC_def}
\end{eqnarray}
The nonconvex function $\phi_{\mathrm{MC}}$ here is known to induce a sparser and less biased 
estimate with respect to the $\ell_1$ penalty. 

\subsection{Primal-dual splitting method}
Let $\mathcal{X}$ and $\mathcal{Y}$ be real Hilbert spaces:
in the present case, $\mathcal{X}:=\mathbb{R}^{N}$ and $\mathcal{Y}:=\mathbb{R}^{n\times n}$.
The primal-dual splitting method \cite{primal-dual} solves convex optimization problems in the following form:
\begin{equation}
    \min _{\bm{w} \in \mathcal{X}}[F(\bm{w})+G(\bm{w})+H(L(\bm{w}))],  \label{pd_algorithm}
\end{equation}
where $F\!:\! \mathcal{X}  \rightarrow \mathbb{R}$ is a differentiable convex function with Lipschitz continuous gradient $\nabla F$, $G\in \Gamma_0(\mathcal{X})\!$ and $H\in \Gamma_0(\mathcal{Y})$ are proximable proper lower semicontinuous convex functions, and $L :  \mathcal{X}  \rightarrow  \mathcal{Y}$ is a bounded linear operator with its adjoint operator denoted by $L^*$. 
Here, ``proximable" means that the proximity operator of the function can be computed easily (in a closed form in the present case).
The primal dual splitting method is given in Algorithm \ref{pds_def}.
\begin{algorithm}[t]      
    \caption{Primal-dual splitting method}         
    \label{pds_def}                          
    \begin{algorithmic}                  
    \REQUIRE Initial estimate $\bm{w}_0 \in \mathcal{X}, \bm{V}_0 \in \mathcal{Y}$, tolerance $\epsilon>0$, proximity parameters $\tau>0$ and $\sigma>0$, relaxation parameters $\rho_{k}>0$.
    \WHILE{$\frac{\|\bm{w}_{k+1}-\bm{w}_{k}\|^2_2}{\|\bm{w}_{k}\|^2_2}> \epsilon$}
    \STATE 1. $\tilde{\bm{w}}_{k+1}:=\operatorname{prox}_{\tau G}\left(\bm{w}_{k}-\tau\nabla F\left(\bm{w}_{k}\right) - \tau L^{*} \bm{V}_{k}\right)$
    \STATE 2. $\tilde{\bm{V}}_{k+1}:=\operatorname{prox}_{\sigma H^{*}}\left(\bm{V}_{k}+\sigma L\left(2 \tilde{\bm{w}}_{k+1}-\bm{w}_{k}\right)\right)$
    \STATE 3. $\left(\bm{w}_{k+1}, \bm{V}_{k+1}\right)\!:=\!\rho_{k}\!\left(\!\tilde{\bm{w}}_{k+1}, \tilde{\bm{V}}_{k+1}\!\right)\!+\!\left(1\!-\!\rho_{k}\right)\!\left(\bm{w}_{k}, \bm{V}_{k}\right)$
    \ENDWHILE
    \end{algorithmic}
\end{algorithm}

\section{PROPOSED ALGORITHMS}
 \label{proposed_algorithms}

Due to its structure (i.e., symmetry and zero row-sums), the CGL
is completely defined by its upper (or lower) triangular part excluding the main diagonal, or, in other words, by a length-$\frac{n(n-1)}{2}$ vector, where the CGL is of size $n\times n$.
Since all the off-diagonal components of CGL need to be nonnegative, 
our variable vector is constrained to the nonnegative cone
(the nonnegative orthant)
$C:= \mathbb{R}_+^\frac{n(n-1)}{2}$.
Given this, we 
define a specific linear operator $L:\! C\!\rightarrow\!\mathbb{R}^{n\times n}$ 
that maps a nonnegative vector of size $n(n-1)/2$ to its corresponding CGL.
For $n=4$, for instance, $L$ is defined as follows:
\setlength{\arraycolsep}{0.35em}
\begin{align*}
    & L: [w_1,w_2,w_3,w_4,w_5,w_6]^T\mapsto\\
  &\hspace{-0.5em}\left[
  \begin{array}{cccc}
  \!w_1\!+\!w_2\!+\!w_3\!   &  -w_1  &   -w_2  & -w_3 \\
  -w_1   &  \!w_1\!+\!w_4\!+\!w_5\!    &  -w_ 4  &-w_5\\
  -w_2   & -w_4 & \!w_2\!+\!w_4\!+\!w_6\! &-w_6\\
 - w_3   & - w_5    &  -w_6   & \!w_3\!+\!w_5\!+\!w_6\! \\
    \end{array}
  \right].
\end{align*}
\setlength{\arraycolsep}{0.5em}
\vspace{-2em}
\subsection{Problem formulation}
The CGL formulation presented in \cite{ortega16} is a popular extension of graphical lasso for imposing a Laplacian constraint.
With a slight modification using the linear operator $L$ introduced above,
the CGL formulation is given by
\begin{align}
    \mathrm{P}_0: \min_{\bm{w}\in C}\; -\logdet(L(\bm{w})\!+\!\bm{J})\!+\!\langle \bm{S}, L(\bm{w}) \rangle \!+\!\lambda_1\|
    \bm{w}\|_{1}, 
\end{align}
where $\bm{J}:=\frac{1}{n}\bm{11}^T\in\mathbb{R}^{n\times n}$, $\bm{S}\in\mathbb{R}^{n\times n}$ stands for the sample covariance obtained from data, and $\lambda_1\geq 0$ is the regularization parameter.
Note here that $L(\bm{w})+\bm{J}$ is positive definite if and only if the graph of $L(\bm{w})$ is connected, i.e., $L(\bm{w})\circ \bm{I} -L(\bm{w})$ is an irreducible matrix, where
$\circ$ denotes the Hadamard product.

Replacing the regularization term of Problem $P_0$ by  the MC penalty given in (\ref{MC_def}), the problem reads as follows:\footnote{Although the formulation in P$_1$ has been considered in the literature \cite{palomar_MC, zhang2020learning}, it was also considered earlier in the authors' previous works \cite{koyakumaru_bthesis, koyakumaru_21} as an intermediate step, and the present study is independent from \cite{palomar_MC, zhang2020learning}.}
\begin{equation}
    \begin{aligned}
	\mathrm{P}_1: \min_{\bm{w}\in C}\; -\logdet(L(\bm{w})+\bm{J})+\langle\bm{S}, L(\bm{w})\rangle \\
	+\lambda_1\underbrace{\left[\|\boldsymbol{w}\|_{1}-^\gamma \!\|\cdot\|_1(\bm{w}) \right].}_{\mathrm{MC}} \label{MCP_p1}
    \end{aligned}
\end{equation}
We introduce the Tikhonov regularization term $\frac{\lambda_2}{2}\|\bm{w}\|_2^2$, 
$\lambda_2\geq 0$,
which also plays a role of convexification as shown below.
    Introducing the indicator function $\iota_C (\bm{w})$ to accommodate the constraint as well, Problem $\mathrm{P}_1$ is transformed into the following unconstrained optimization problem:
            \begin{align} \mathrm{P_2} :  
         &\min_{\bm{w}\in \mathbb{R}^{\frac{n(n-1)}{2}}}\;
                \underbrace{-\lambda_1 \ ^\gamma \|\cdot\|_{1}( \boldsymbol{w})+\frac{\lambda_2}{2}\|\bm{w}\|_2^2}_{F(\bm{w})}
    	\nonumber\\
&        +\underbrace{\iota_C(\bm{w})\!+\!\lambda_1\|\bm{w}\|_1\!+\!\langle \bm{S}, L(\bm{w}) \rangle }_{G(\bm{w})} \underbrace{-\logdet (L(\bm{w})\!+\!\bm{J})}_{H (L(\bm{w}))}. \label{cost_fn_detail}
        \end{align}
         By using Moreau's decomposition $\frac{1}{2\gamma}\|\cdot\|_2^2 =\hspace{0.1em} ^\gamma f +\hspace{0.1em} ^{\gamma^{-1}}f^{*}  \circ \gamma^{-1} \, \bm{I}$ \cite[Theorem 14.3]{combettes_book}, the MC penalty term can be rewritten as
\begin{eqnarray}
\hspace{-2em}
-^\gamma \|\cdot\|_1(\bm{w}) &\!=\!& ^{\gamma^{-1}}(\| \cdot \|_1^*) (\gamma^{-1} \bm{w})\!-\!\frac{1}{2\gamma}\|\bm{w}\|_2^2 \\
&\!=\!& ^{\gamma} \iota_{B_\infty}(\bm{w}) - \frac{1}{2\gamma}\|\bm{w}\|_2^2, \label{eq:l1_conjugate}
\end{eqnarray}
    where $B_\infty  := \mathrm{lev}_{\leq \gamma}\| \cdot \|_\infty := \{\bm{x} \in \mathbb{R}^{n} \mid \|\bm{x}\|_\infty \leq \gamma \}$ is  the $\ell_\infty$ ball of radius-$\gamma$ \cite[Example 13.32]{combettes_book}.
    The usefulness of this decomposition of the MC penalty term has been observed also in \cite{suzuki21, kaneko20,yukawa21_alime, komuro_2021_SSP}.
    Using \eqref{eq:l1_conjugate}, the function $F$ of P$_2$ can be rewritten as
            \begin{equation}
        F(\boldsymbol{w})= \lambda_1 \ ^{\gamma} \iota_{B_\infty}(\bm{w})
    			-\frac{\lambda_1}{2\gamma}\|\bm{w}\|_2^2 + 
    			\frac{\lambda_2}{2}\|\bm{w}\|_2^2,
    			\label{eq:Fw}
    \end{equation}
    of which the convexity is ensured clearly by choosing $\lambda_1$ and $\lambda_2$ such that
    $\lambda_2\geq \gamma^{-1} \lambda_1$ (see Proposition \ref{convergence_condition} below). 
    The Tikhonov regularization term $\frac{\lambda_2}{2}\|\bm{w}\|_2^2$ thus has a convexification property, as mentioned above.
On the other hand, the functions $G(\boldsymbol{w})$ and $H\circ L(\boldsymbol{w})$ are convex,
since the composition of a convex function with an arbitrary affine operator is also a convex function.
Hence, under the convexity condition given above,
Problem $\mathrm{P_2}$ takes the form of \eqref{pd_algorithm},
and it can be solved by the primal-dual splitting method.

\subsection{Optimization algorithm}
 The proposed algorithm is derived by applying the primal-dual spitting method to Problem $\mathrm{P_2}$. 
 \subsubsection{Derivation of $\tilde{\bm{w}}_{k+1}$}
We define the soft thresholding operator
for a length-$\frac{n(n-1)}{2}$ positive vector $\bm{\delta}:=[\delta_1,\delta_2,\cdots,\delta_\frac{n(n-1)}{2}]^{T}$ by
  \begin{eqnarray}
        [\mathrm{soft}_{\bm{\delta}}(\bm{w})]_i=\left\{
            \begin{array}{lll}
                w_i-\delta_i, &\mathrm{if}	~ w_i \geq \delta_i,\\
                0,         &\mathrm{if}~ |w_i|<\delta_i,   \\
                w_i+\delta_i, &\mathrm{if}	~ w_i \leq -\delta_i,\\
            \end{array}
            \right.
    \end{eqnarray}
    where $[\cdot]_i$ is the $i$th component of the argument.
   The convex projection onto the nonnegative cone $C$ is given by
    \begin{eqnarray}
        [P_{C}(\bm{w})]_i =\left\{
            \begin{array}{lll}
                w_i , &\mathrm{if}	~ w_i \geq 0,\\
                0,         &\mathrm{if}~ w_i < 0.   \\
            \end{array}
            \right.
    \end{eqnarray}
   Applying Step 1 of Algorithm \ref{pds_def} to Problem $\mathrm{P_2}$ yields
  \begin{align}
        \tilde{\bm{w}}_{k+1}&=\operatorname{prox}_{\tau G}
        \bigl[
            \bm{w}_{k}-\tau L^{*} (\bm{V}_{k})-\tau\nabla F\left(\bm{\bm{w}}_{k}\right)
        \bigr]\\ 
        &= \operatorname{prox}_{\tau G}
        \Bigl[\bm{w}_{k}-\tau L^{*} (\bm{V}_{k})-\tau\bigl(\gamma^{-1}_{}\lambda_1 \prox_{\|\cdot\|_1}(\bm{w}_k) \nonumber \\
            &\quad-\gamma^{-1}_{}\lambda_1\bm{w}_k+\lambda_2\bm{w}_k \bigr)\Bigr].\
            \label{before_prop1}
    \end{align}
    The operators $L^*$ and $\prox_{\tau G}$ can be computed by using
    the following  propositions.
\begin{proposition}
    \label{PDS_DEF_ADJOINT}
    \label{pds_def_adjoint}
Let $\bm{M}\in\mathbb{R}^{n\times n}$ be an arbitrary CGL matrix with its
$(p,q)$ component denoted by $m_{p,q}$.
Then, for any $p,q\in\{1,2,\cdots,n\}$ such that
$(2n-p-1)p/2+q-n\in\{1,2,\cdots,n(n-1)/2\}$,
    it holds that
    \begin{equation*}
    [L^*(\bm{M})]_{(2n-p-1)p/2+q-n}=m_{p,p}+m_{q,q}-m_{p,q}-m_{q,p}.
    \end{equation*}
\end{proposition}
\noindent {\bf  Proof:} See Appendix \ref{appendix_1}.
\begin{proposition}
\label{PROX_INNER_PROD}
\label{prox_inner_prod}
The proximity operator of $G(\bm{w}) \! := \! \iota_c(\bm{w}) \! + \! \lambda_1\|\bm{w}\|_1 \! + \! \langle \bm{S}, L(\bm{w}) \rangle$ 
of index $\tau>0$ can be expressed by 
\begin{eqnarray}
    \prox_{\tau G}(\bm{w}) = P_C(\bm{w}-\tau(\lambda_1 \bm{1}+ L^*(\bm{S}))).
\end{eqnarray}
\end{proposition}
\noindent {\bf  Proof:} See Appendix  \ref{appendix_2}.

By using Proposition \ref{prox_inner_prod} and $\mathrm{prox}_{\|\cdot\|_1} = \mathrm{soft}_{\bm{1}}$, \eqref{before_prop1} can be 
rewritten as
\begin{eqnarray}
   \tilde{\bm{w}}_{k+1}=P_{C}
   [
        \bm{w}_{k}-\tau L^{*}( \bm{V}_{k})-\tau(\lambda_1\bm{1}+L^*(\bm{S})) \nonumber\\ 
        -\tau(\gamma^{-1}\lambda_1 \soft_{\bm{1}}(\bm{w}_k)-\gamma^{-1}\lambda_1\bm{w}_k+\lambda_2\bm{w}_k)
    ].
    \label{eq:def_wtilde}
\end{eqnarray}

\subsubsection{Derivation of $\tilde{\bm{V}}_{k+1}$}

Substituting Moreau's decomposition \cite[Theorem 14.3]{combettes_book}
\begin{eqnarray}
        \operatorname{prox}_{\sigma H^{*}}(\bm{u})&=&\bm{u}-\sigma\operatorname{prox}_{\sigma^{-1}H}(\sigma ^{-1}\bm{u}) \label{prox_dual}
    \end{eqnarray}
 with $H:=-\logdet (\cdot +\bm{J})$
into Step 2 of Algorithm \ref{pds_def} yields
\begin{align}
\tilde{\bm{V}}_{k+1}&\!=\!\bm{V}_{k}+\sigma L\left(2 \tilde{\bm{w}}_{k+1}-\bm{\bm{w}}_{k}\right) \nonumber\\
 -&\sigma\operatorname{prox}_{\sigma^{-1} \left(-\logdet(\cdot +\bm{J})\right)}\left[\sigma^{-1}\bm{\bm{V}}_{k}\!+\!L\left(2 \tilde{\bm{w}}_{k+1}\!-\!\bm{\bm{w}}_{k}\right)\right]. \nonumber \\
\label{y_np1 op}
\end{align}
Here, the proximity operator $\prox_{\sigma^{-1} (-\logdet(\hspace*{.2em}\cdot \hspace*{.2em}+\hspace*{.2em}\bm{J}))}$ 
can be written in a closed form, as shown in the following proposition.

\begin{proposition}
\label{PROP_PROX_LOGDET}
\label{prop_prox_logdet}
For a positive semi-definite matrix $\bm{W} \in \mathbb{R}^{n\times n}$, it holds that 
    \begin{align}
        &\prox_{\sigma^{-1}(-\logdet(\hspace*{.2em}\cdot \hspace*{.2em}+\hspace*{.2em}\bm{J}))}(\bm{W}) \nonumber \\
        &=\bm{Q}\,\diag\left(\!\frac{\mu_{1}\!+\!\sqrt{\mu_1^2\!+\!4\sigma^{-1}}}{2},\! \cdots \!, \frac{\mu_{n}\!+\!\sqrt{\mu_n^2\!+\!4\sigma^{-1}}}{2}\right)\!\bm{Q}^T \nonumber\\
        &\quad-\bm{J}, 
    \end{align}
where $\mu_{i}$ is the $i$th eigenvalue of $\bm{W}+\bm{J}$, and $\bm{Q}:=[\bm{q}_1 \bm{q}_2 \ldots \bm{q}_n]$ with the eigenvectors $\bm{q}_i$ of $\bm{W}+\bm{J}$. \\
\noindent {\bf  Proof:} See Appendix  \ref{appendix_3}.
\end{proposition}

\noindent An application of Proposition \ref{prop_prox_logdet} to \eqref{y_np1 op} yields
\begin{align}
&\tilde{\bm{V}}_{k+1}\!=\!\bm{V}_{k}+\sigma L\left(2 \tilde{\bm{w}}_{k+1}-\bm{w}_{k}\right)+\sigma\bm{J}\nonumber\\ &-\!\sigma\!\left[\!\bm{U}\diag\!\left(\!\dfrac{\nu_{1}\!+\!\sqrt{\nu_1^2\!+\!4\sigma^{-\!1}}}{2}\!,\! \cdots \! ,\! \dfrac{\nu_{n}\!+\!\sqrt{\nu_n^2\!+\!4\sigma^{-\!1}}}{2}\!\right)\!\bm{U}^T\!\right] ,
\label{eq:def_Vtilde}
\end{align}
where $\nu_{i}$ is the eigenvalue
of $\sigma^{-1}\bm{\bm{V}}_{k}+L\left(2 \tilde{\bm{w}}_{k+1}-\bm{\bm{w}}_{k}\right)$, and $\bm{U}:=[\bm{u}_1 \bm{u}_2 \ldots \bm{u}_n]$ with the eigenvectors $\bm{u}_i$ of $\sigma^{-1}\bm{\bm{V}}_{k}+L\left(2 \tilde{\bm{w}}_{k+1}-\bm{\bm{w}}_{k}\right)$.
\\
The proposed algorithm is given in Algorithm \ref{alg2}.
Our formulation based on the MC penalty is expected to yield 
a sparser solution with better interpretability than the conventional $\ell_1$-based methods due to the efficient sparsity promoting property of the MC penalty.  In addition, our representation of CGL using the linear operator reduces the number of variables approximately by half, while also transforming the positive semi-definite constraint of the  graph Laplacian to the nonnegativity  constraint $\bm{w} \in C$. However,  our formulation needs $\mathcal{O}(n^3)$ complexity due to the need to execute matrix multiplication and eigenvalue decomposition. This computational drawback can be mitigated by using the eigenvalue decomposition method for symmetric matrices \cite{MagoarouGT16}. Note that, in the particular case of $\gamma:= +\infty$, the proposed algorithm gives an alternative way to solve  the graphical lasso problem for CGL.

\begin{algorithm}[t]                      
    \caption{Proposed graph learning algorithm}         
    \label{alg2}                          
    \begin{algorithmic}                  
    \REQUIRE Initial estimate ($\bm{w}_0, \bm{V}_0$), tolerance $\epsilon>0$, proximity parameters $\tau>0$, $\sigma>0$, covariance matrix $\bm{S}$, regularization parameter $\lambda_1 \geq 0, \lambda_2 \geq 0$, minimax concave parameter $\gamma^{-1} >0$, relaxation parameters $\rho_k >0.$
    \\
    \ENSURE Graph Laplacian $\bm{\Theta}$    
    \WHILE{$\frac{\|\bm{w}_{k+1}-\bm{w}_{k}\|^2_2}{\|\bm{w}_{k}\|^2_2}> \epsilon$}
     \STATE 1. Compute the vector $\tilde{\bm{w}}_{k+1}$ by \eqref{eq:def_wtilde}
    \STATE 2.~Find the eigenvalues $\nu_i$ and the matrix 
    $\bm{U}=[\bm{u}_1 \ldots  \bm{u}_n]$
    containing all the corresponding (unit-norm) eigenvectors of
        $\left(\bm{J}+\sigma^{-1}\bm{V}_{k}+L \left(2 \tilde{\bm{w}}_{k+1}-\bm{w}_{k} \right) \right)$
    \STATE 3. Compute the vector $\tilde{\bm{V}}_{k+1}$ by \eqref{eq:def_Vtilde}
    \STATE 4. $\left(\bm{w}_{k+1}, \bm{V}_{k+1}\right) \!=\!\rho_k \!\left(\tilde{\bm{w}}_{k+1}, \tilde{\bm{V}}_{k+1}\right)\!+\!\left(1\!-\!\rho_k \right)\!\left(\bm{w}_{k}, \bm{V}_{k}\right)$
    \ENDWHILE
    \RETURN $\bm{\Theta}=L(\bm{w}_k)$
    \end{algorithmic}
\end{algorithm}
\subsection{Convergence conditions}
\label{sec_convergence_condition}
Convergence is guaranteed under the following conditions. 
\begin{proposition}
\label{CONVERGENCE_CONDITION}
\label{convergence_condition}
If $\lambda_2 \geq \gamma^{-1} \lambda_1$, the function $F$ is convex.
In this case, 
Algorithm 1 converges to a global minimizer of $\mathrm{P}_2$ if 
the following conditions are jointly satisfied:
\begin{enumerate}
    \item $\dfrac{1}{\tau} \geq 2\sigma n+\dfrac{\lambda_2}{2},$
    \item $0 < \rho_k < 2-\dfrac{\lambda_2}{2}\left(\dfrac{1}{\tau}-2\sigma n\right)^{-1}.$
    
\end{enumerate}
\end{proposition}
\noindent {\bf Proof: } See Appendix  \ref{appendix_4}.

From a theoretical side,
a use of $\lambda_2$ satisfying the convexity condition shown in Proposition \ref{convergence_condition}
ensures convergence to a global minimizer.
From a practical side, on the other hand,
a use of $\lambda_2$ violating the convexity condition may yield better performance,
as will be seen in Section \ref{numerical_example}.
However, we emphasize that improved  performance comes with 
no theoretical guarantees.
A remarkable advantage of the present framework is 
its flexibility due to the use of powerful convex analytic solver, which  allows to extend the presented framework 
in many possible directions including
dynamic graph learning \cite{HallacPBL17,tanaka19_time_varying}.

\section{Numerical Examples}
\label{numerical_example}
We show the efficacy of the proposed method through some experiments with synthetic and real data.
We first show the performances of the proposed method for different regularization parameters.
We then compare the performance of the proposed method with CGL estimation \cite{ortega16} and SGL estimation \cite{ palomar20}.
\footnote{We used the implementations of the SGL and CGL algorithms in {\tt spectralGraphTopology} (\url{https://CRAN.R-project.org/package=spectralGraphTopology}).}

\subsection{Experiments with synthetic data}
\label{Comparisons_estimate_accuracy}
\noindent{\bf Dataset generation:} 
We consider three types of graph:
(i) grid graph $\mathcal{G}_{\mathrm{grid}}^{(\sqrt{n}, \sqrt{n})}$  with nodes connected to their
four nearest neighbors (except the nodes at boundaries),
(ii) random modular graph (a.k.a.~stochastic block model) $\mathcal{G}_{\mathrm{M}}^{(n, 0.01, 0.3)}$ with four modules where the nodes are connected 
across the modules and within each module with probabilities 0.01 and 0.3, respectively, and (iii) Erd\"os-R\'{e}nyi graph $\mathcal{G}_{\mathrm{ER}}^{(n, 0,1)}$ with nodes connected to other nodes with probability 0.1.
The graph weights are randomly drawn from the uniform distribution over the interval [0.1, 3.0], regarded as 
the ground-truth graph Laplacian $\bm{\Theta}$ in this experiment.
From each graph generated, data are generated from $\mathcal{N}(0, \bm{\Theta}^{\dagger})$, where $(\cdot)^\dagger$ denotes the Moore-Penrose pseudo inverse, and the covariance matrix $\bm{S}$ is computed from data.
For each type of graph,
we randomly generate $15$ graphs with $n=100$ nodes using 
the toolbox given in \cite{perraudin2014gspbox}. 

\noindent{\bf Performance measure:} 
The relative error (RE) and F-score (FS) are used as performance measures:
\begin{gather}
    \operatorname{RE}\left(\widehat{\bm{\Theta}}, \bm{\Theta}_{\star}\right):=\frac{\left\|\widehat{\bm{\Theta}}-\bm{\Theta}_{\star}\right\|^2_{\rm F}}{\left\|\bm{\Theta}_{\star}\right\|^2_{\rm F}},\\
    \mathrm{FS}\left(\widehat{\bm{\Theta}}, \bm{\Theta}_{\star}\right):=\frac{2 \mathrm{tp}}{2 \mathrm{tp}+\mathrm{f} \mathrm{n}+\mathrm{fp}},
\end{gather}
where
$\mathrm{tp, fp}$, and $\mathrm{fn}$ stand for true-positive, false-positive, and false-negative, respectively. 
Here, RE indicates the discrepancy between the ground-truth graph Laplacian $\bm{\Theta}_{\star}\in \mathbb{R}^{n\times n}$
and its estimate $\widehat{\bm{\Theta}}\in\mathbb{R}^{n\times n}$, while
FS is a measure of accuracy of binary classification (taking values in [0,1]), 
indicating whether the sparse structures are extracted correctly.

\subsubsection{Performance of the proposed method}
\label{subsubsec:performance_proposed}
\begin{figure}[t!]

    \begin{minipage}{1.0\linewidth}
        \centering
        \subfloat[RE]{\includegraphics[width=6.5cm,clip]{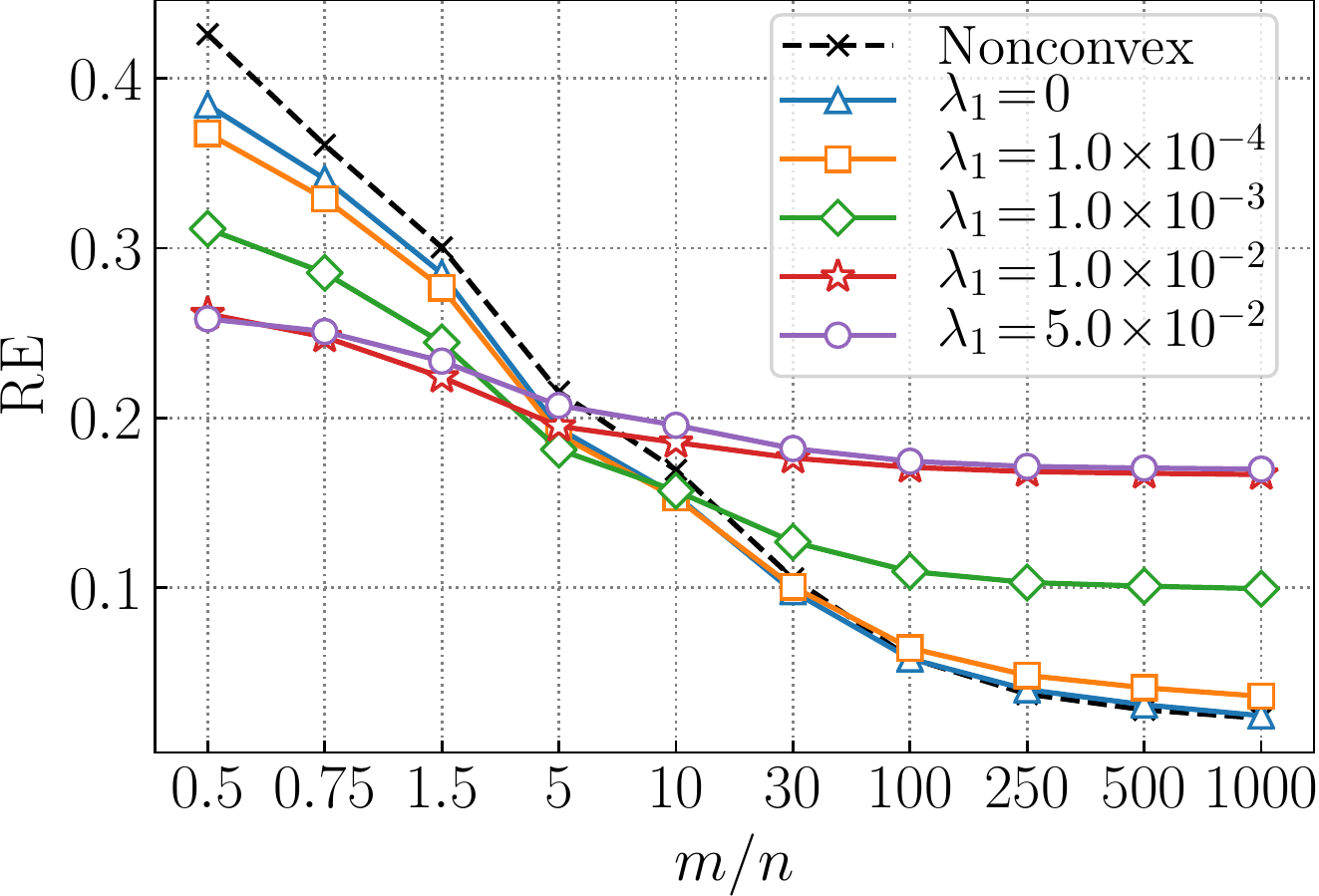}}
    \end{minipage}
    
    \begin{minipage}{1.0\linewidth}
        \centering
        \subfloat[FS]{\includegraphics[width=6.5cm,clip]{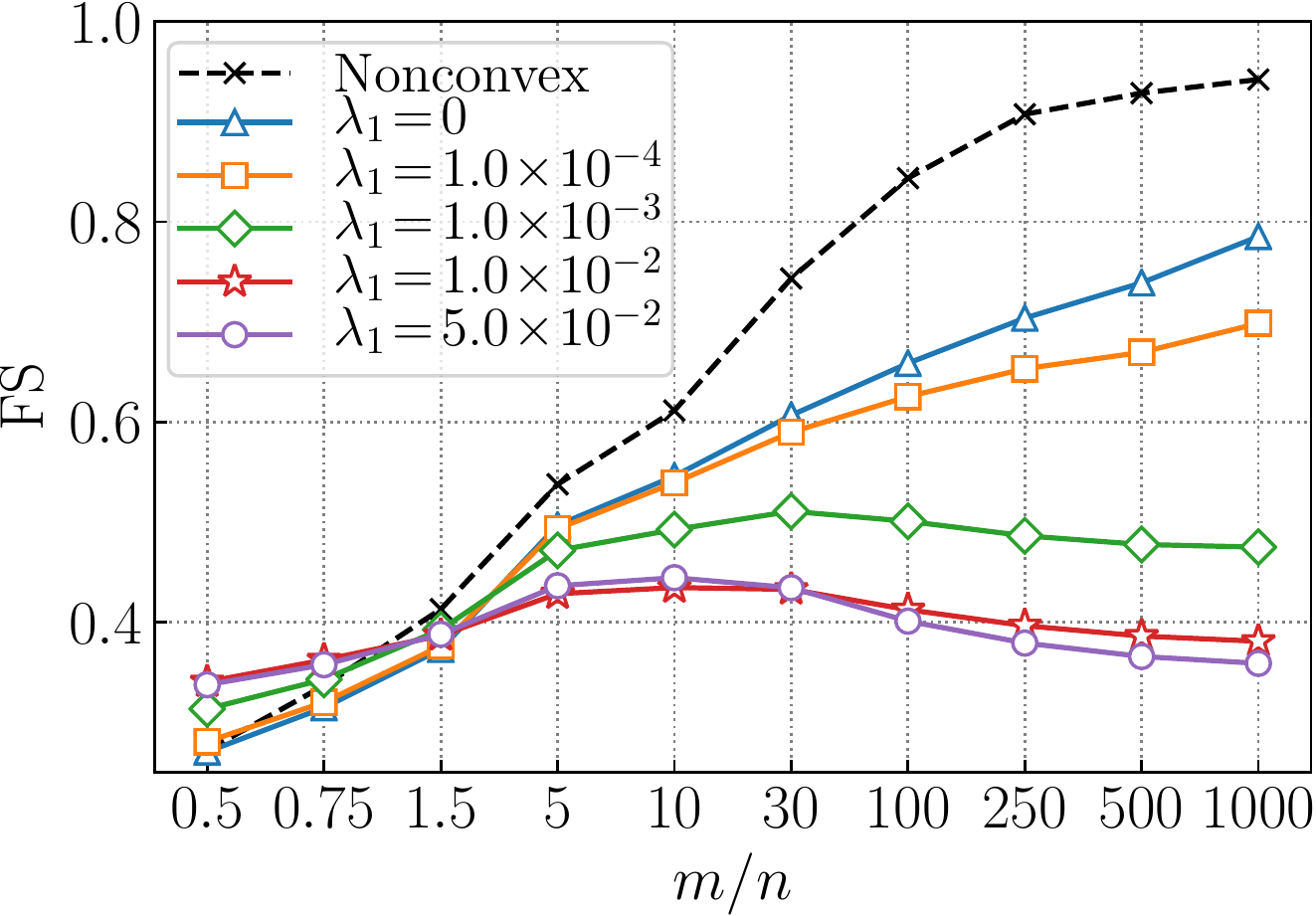}}
    \end{minipage}
        \caption{Experimental results of each $\lambda_1$ value under the convexity condition in random modular graph $\mathcal{G}_{\mathrm{M}}^{(100, 0.01, 0.3)}$ estimation.}
        \label{modular_change_lmd_RE_FS}
\end{figure}

\begin{figure}[t!]

    \begin{minipage}{1.0\linewidth}
        \centering
        \subfloat[RE]{\includegraphics[width=6.5cm,clip]{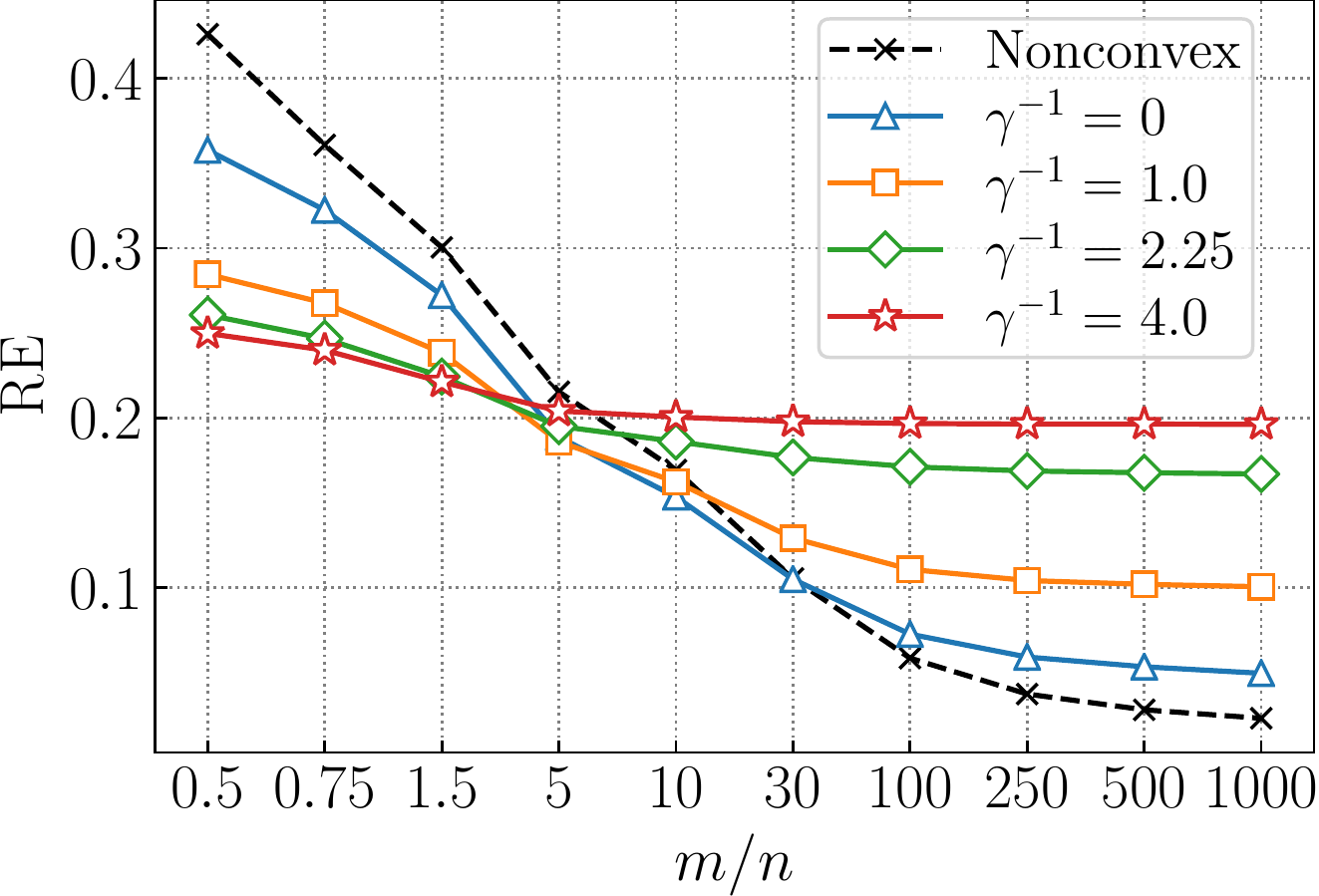}}
    \end{minipage}
    
    \begin{minipage}{1.0\linewidth}
        \centering
        \subfloat[FS]{\includegraphics[width=6.5cm,clip]{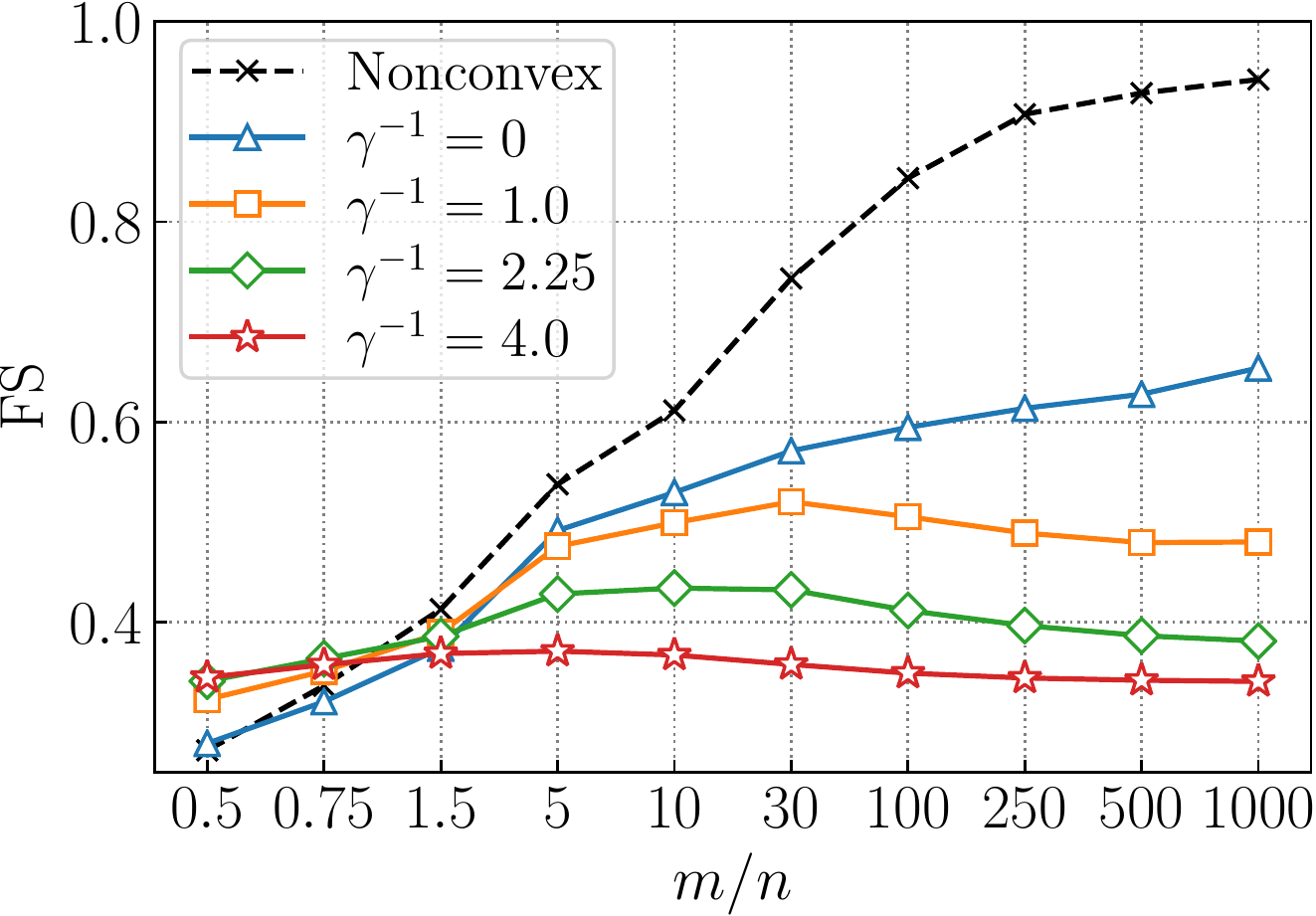}}
    \end{minipage}
        \caption{Experimental results of each $\gamma^{-1}$ value under the convexity condition in random modular graph $\mathcal{G}_{\mathrm{M}}^{(100, 0.01, 0.3)}$ estimation.}
        \label{modular_change_m_RE_FS}
\end{figure}

We study the impacts of the parameters $\lambda_1$ and $\gamma^{-1}$
of the MC penalty
on the performance of the proposed method.
We also tested the case of $\lambda_2:=0$, which makes the entire cost function 
nonconvex for any $\lambda_1>0$, with the other parameters tuned manually (see Table \ref{parameters}).
To study the impact of $\lambda_1$, we fix $\gamma^{-1}:=2.25$ which gave a best performance in the nonconvex case.
Since $\tau$ is the algorithm parameter and it only affects the convergence speed, we fix it to $\tau:=1.0$.
The other parameters are then set to $\lambda_2 := \gamma^{-1}\lambda_1$
and $\sigma\approx(1/\tau-\lambda_2/2)/(2n)$
according to the convexity condition (See Proposition \ref{convergence_condition}).
Figure \ref{modular_change_lmd_RE_FS} plots the RE and FS curves across $m/n$ in modular graph $\mathcal{G}_{\mathrm{M}}^{(100, 0.01, 0.3)}$ for different choices of $\lambda_1$.
To study the impact of $\gamma^{-1}$, on the other hand,
we fix $\lambda_1:= 1.0\times 10^{-4}$ and choose the other parameters in the same way as in Fig.~\ref{modular_change_lmd_RE_FS}.
 Figure \ref{modular_change_m_RE_FS} plots the RE and FS curves for different choices of  $\gamma^{-1}$ under the same conditions as in Fig.~\ref{modular_change_lmd_RE_FS}, with $\lambda_1$=$1.0 \times 10^{-4}$.
 In Figs.~\ref{modular_change_lmd_RE_FS} and \ref{modular_change_m_RE_FS}, the proposed method using the convexity condition attains better performance
 than the nonconvex case when $m/n$ is small, while the nonconvex case is better when  $m/n$ is large.
 To be more specific, when $m/n$ is small, using larger $\lambda_1$, or larger $\gamma^{-1}$, yields better performance.
 Although the regularization parameter $\lambda_2$ for the Tikhonov regularization
needs to be sufficiently large  to ensure the convexity of the entire objective,
using a $\lambda_2$ that is too large tends to yield a less sparse solution, which means
degradation of graph interpretability (cf.~Section \ref{sec_convergence_condition}).

\renewcommand{\arraystretch}{0.6}
\begin{table}[t!]
    \caption{Parameter settings for each graph.}
    \label{parameters}
    \centering
    \small
    \begingroup
    \renewcommand{\arraystretch}{1.2}
    \begin{tabular}{|c|c|c|c|c|}
      \hline
       \multicolumn{2}{|c|}{parameter}  & grid & modular & ER  \\
      \hline \hline
      &$\lambda_1$ &0.005&0.01&0.01\\
     \raisebox{-0.5em}[1em][0em]{nonconvex} &$\lambda_2$  & 0&0&0\\
     &$\tau$  &1.0&1.0& 1.0\\
      &$\sigma$  & 0.05&0.05&0.01 \\ \hline
      &$\lambda_1$  &$1.0\!\times\! 10^{-4}$&$1.0\!\times\! 10^{-4}$ & $1.0\!\times\! 10^{-4}$ \\
    \raisebox{-0.5em}[1em][0em]{convex} &$\lambda_2$ &$2.5\!\times\! 10^{-4}$&$2.5\!\times\! 10^{-4}$ & $2.5\!\times\! 10^{-4}$\\
    &$\tau$  &1.0&1.0& 1.0\\
      &$\sigma$ & $4.9\!\times\! 10^{-3}$ &$4.9\!\times\! 10^{-3}$&$4.9\!\times\! 10^{-3}$ \\
      \hline
      \multicolumn{2}{|c|}{$\gamma^{-1}$}&\multicolumn{3}{|c|}{2.25}\\
      \multicolumn{2}{|c|}{$\rho_k$} &\multicolumn{3}{|c|}{1.0}\\
      \multicolumn{2}{|c|}{Maximum iterations}&\multicolumn{3}{|c|}{5000}\\
      \multicolumn{2}{|c|}{Tolerance error} &\multicolumn{3}{|c|}{$1.0\times10^{-4}$}\\
      \hline
    \end{tabular}
    \endgroup
  \end{table}
  
\subsubsection{Comparisons in estimation accuracy}

\noindent{\bf Parameters:} 
The best parameters are chosen manually, see Table \ref{parameters}.
The generated graphs depend only on $\lambda_1$, $\lambda_2$, and $\gamma^{-1}$,
while being independent of the algorithm parameters $\tau$, $\sigma$, and $\rho_k$ in principle.
It is very important to tune the parameters $\lambda_1$, $\lambda_2$, and $\gamma^{-1}$ carefully
for better performance, although the generated graph changes gradually as each of those parameters changes.
As shown in Section \ref{subsubsec:performance_proposed}, for the proposed method under the convexity condition, $\lambda_1 =1.0 \times 10^{-4}$ is used.
We mention that a smaller value of $\lambda_1$ tends to give a better result for RE and FS 
when $m/n$ is large. Although the convergence is guaranteed under the condition of $\lambda_2$ described in Section \ref{sec_convergence_condition}, $\lambda_2 := 0$ (for which there is no guarantee of convergence to the global minimizer) 
gave the best performance in the current experiments. 
Regarding the parameters for CGL and SGL, we follow the parameter selection techniques proposed in \cite{ortega16} and \cite{palomar20}, respectively. 

\begin{figure}[t!]
 \hspace{-2em}
        \begin{tabular}{l}    
    \begin{minipage}[htb]{0.5\linewidth}
        \centering
        \subfloat[Ground truth]{\includegraphics[width=4.3cm,clip]{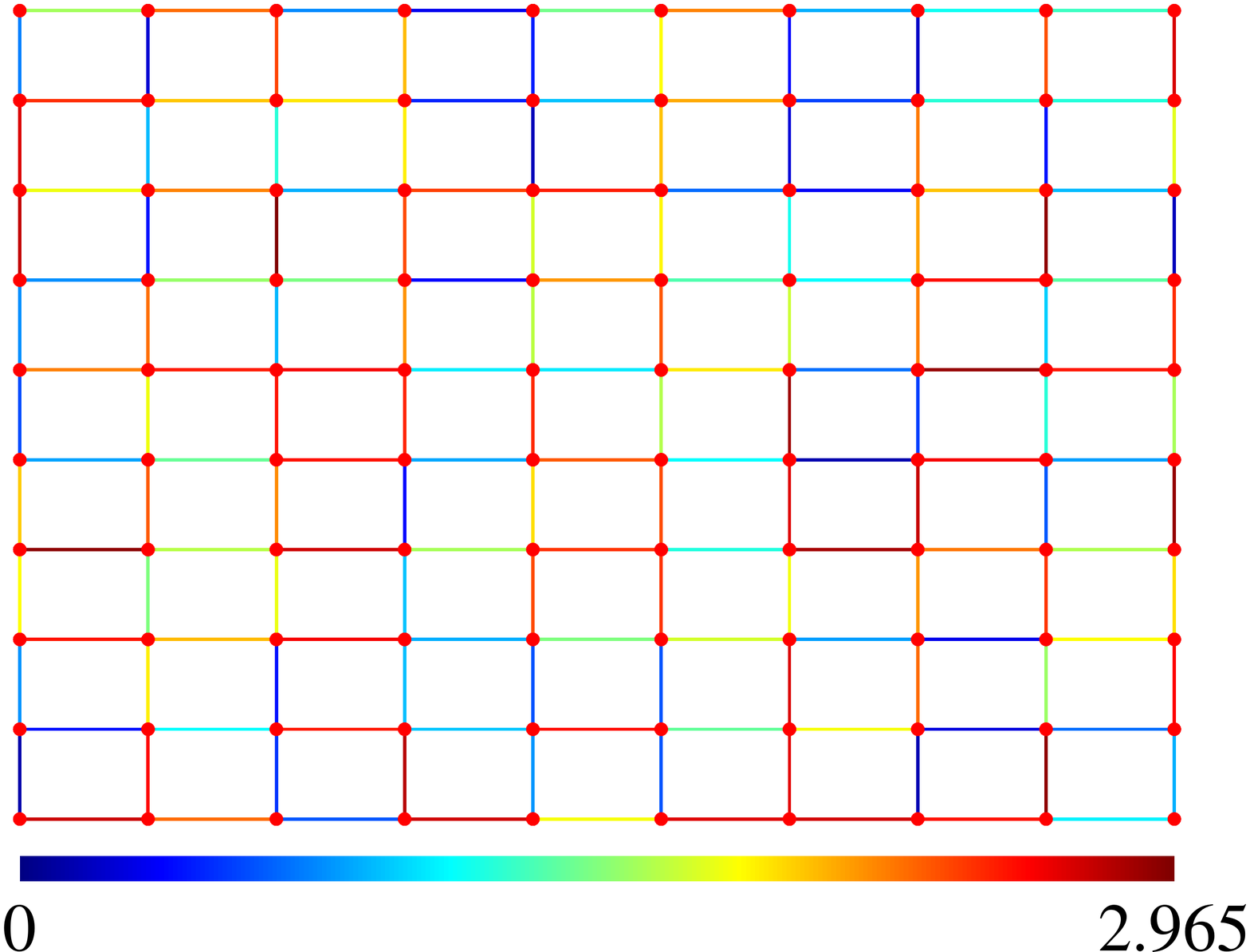}}
    \end{minipage}

    \begin{minipage}[htb]{0.5\linewidth}
        \centering
        \subfloat[Proposed]{\includegraphics[width=4.3cm,clip]{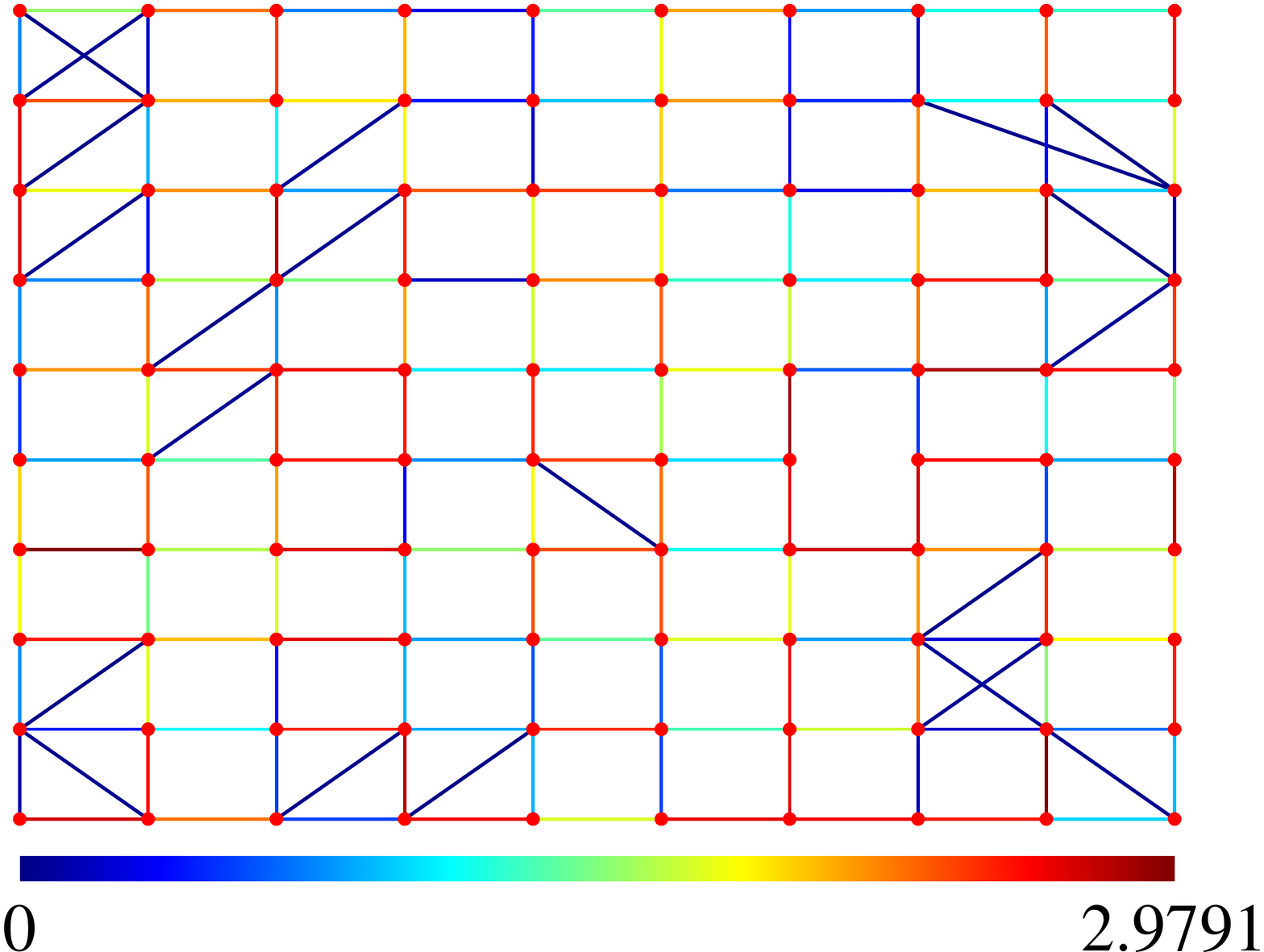}}
    \end{minipage}
    \end{tabular}
    
     \hspace{-2em}
    \begin{tabular}{l}    
        \begin{minipage}[htb]{0.5\linewidth}
            \centering
            \subfloat[CGL]{\includegraphics[width=4.3cm,clip]{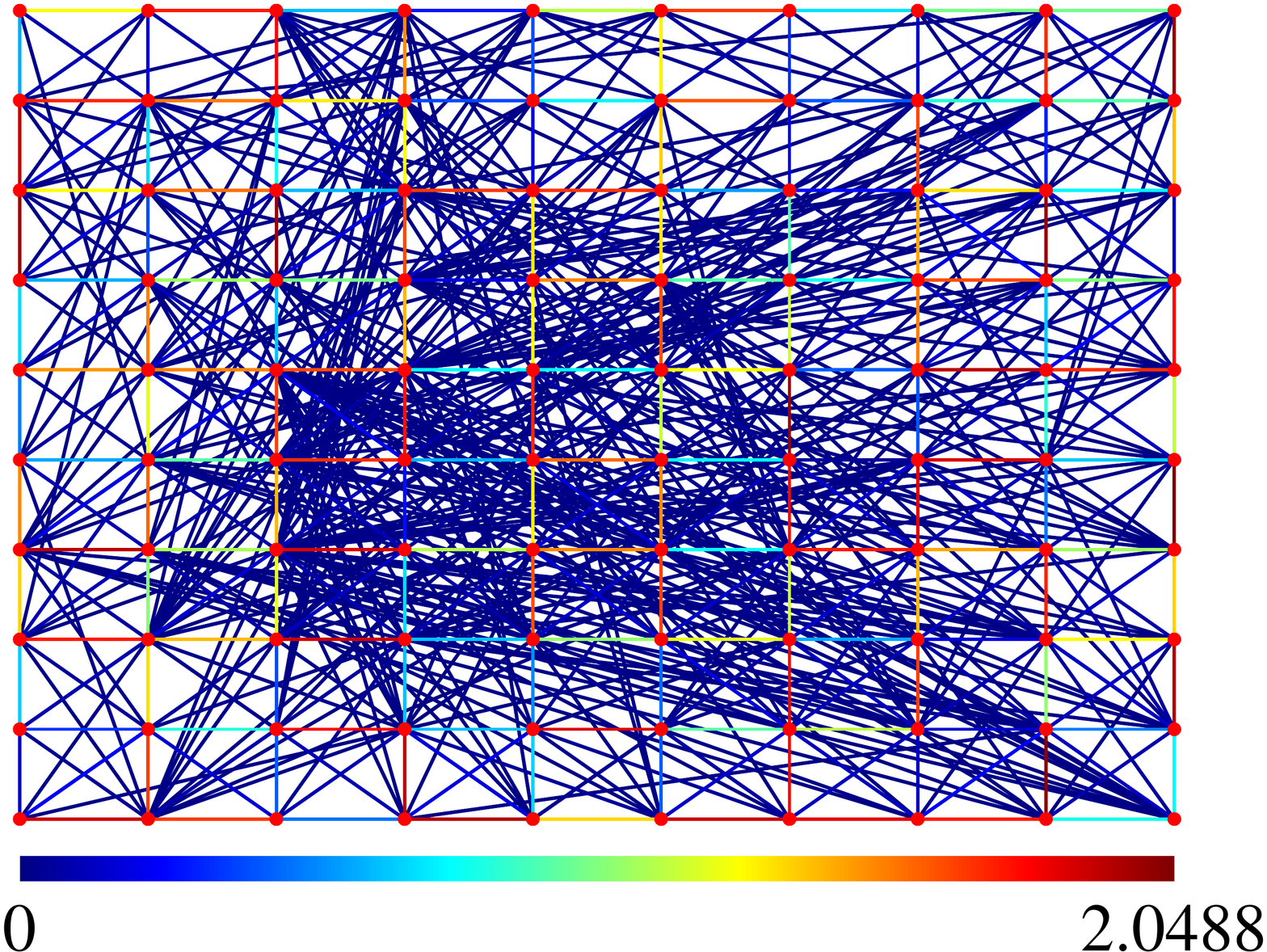}}
        \end{minipage}
    
        \begin{minipage}[htb]{0.5\linewidth}
            \centering
            \subfloat[SGL]{
            \includegraphics[width=4.3cm,clip]{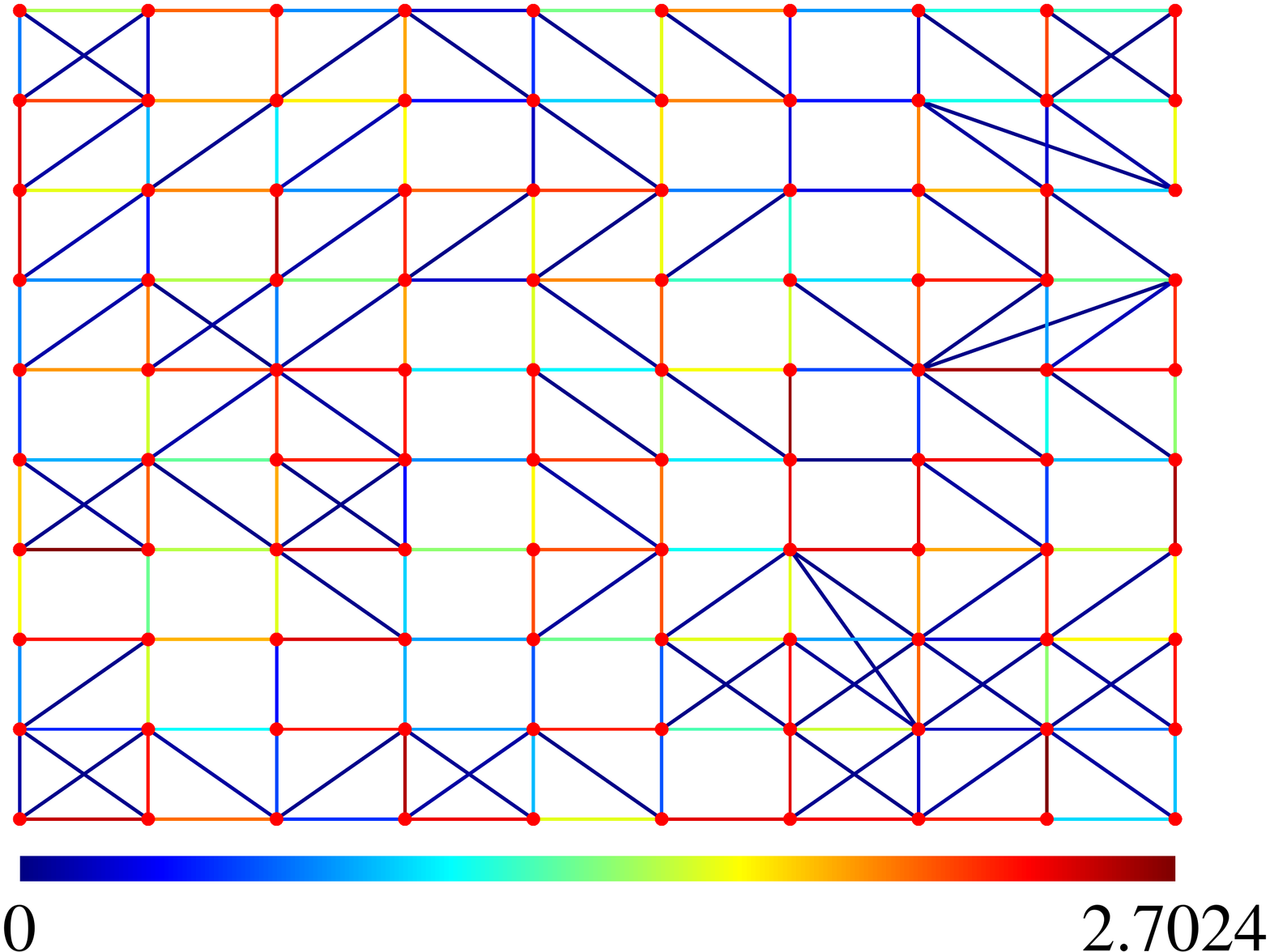}}
        \end{minipage}
    \end{tabular}
    \caption{Visualization of estimate graphs, in which the edge colors
    represent the edge weights.}
    \label{graph_visualize}
    \end{figure}

\begin{figure}[t!]
\begin{minipage}{1.0\linewidth}
    \centering
    \subfloat[RE]{\includegraphics[width=6.5cm,clip]{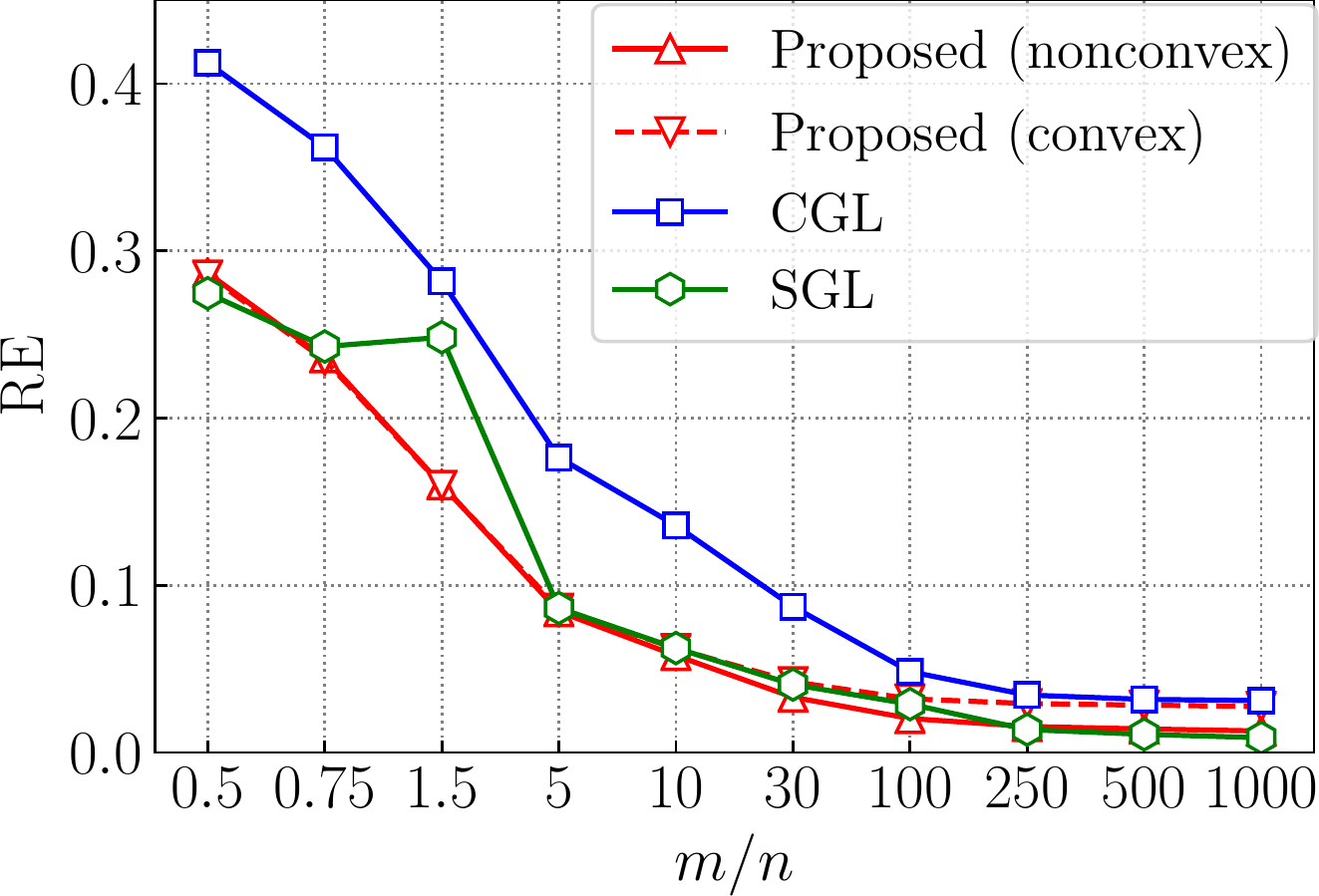}}
\end{minipage}

\begin{minipage}{1.0\linewidth}
    \centering
    \subfloat[FS]{\includegraphics[width=6.5cm,clip]{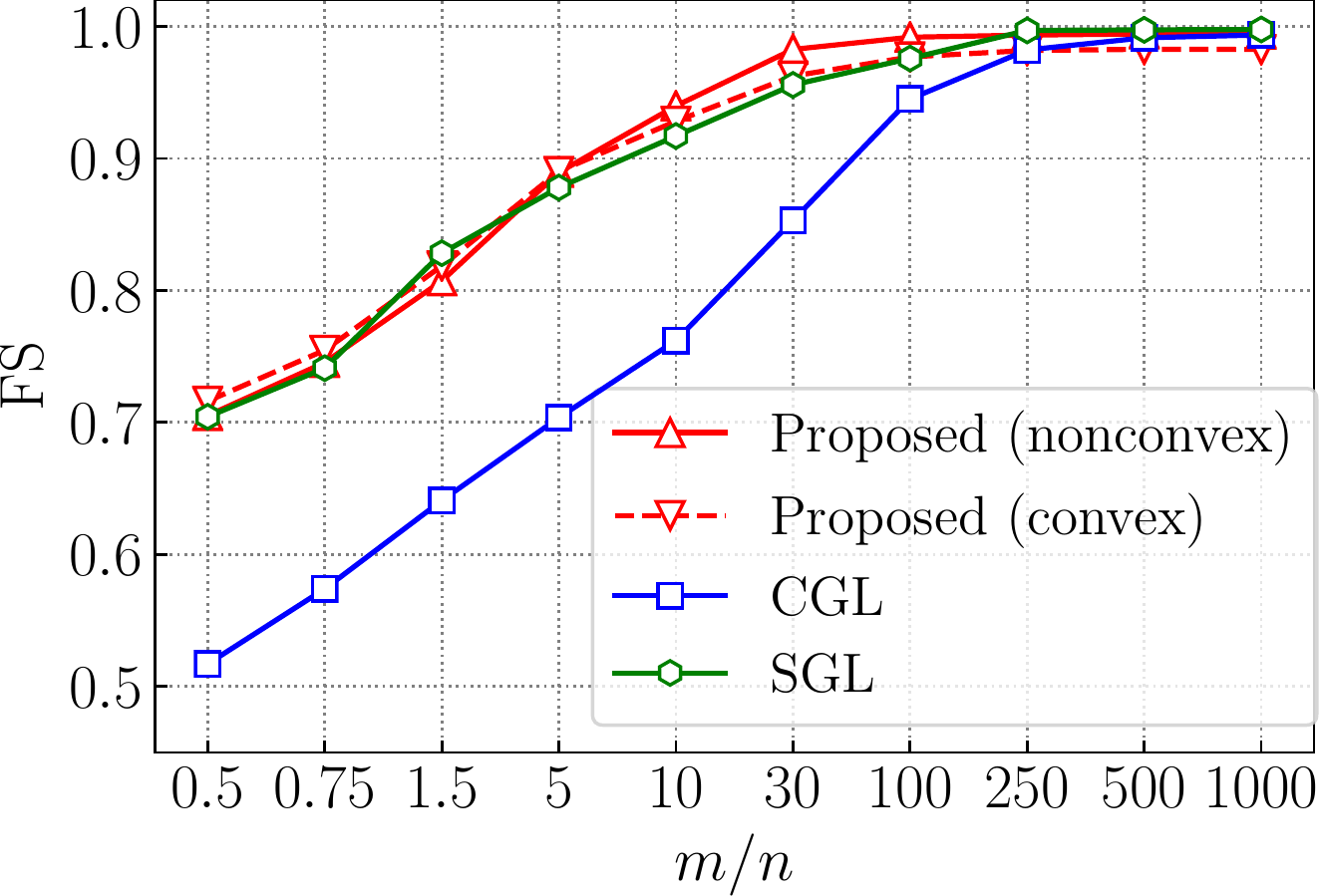}}
\end{minipage}
\caption{Experimental results for grid graph $\mathcal{G}_{\mathrm{grid}.}^{(10, 10)}$}
\label{grid_RE_FS}
\end{figure}

\begin{figure}[t!]
    \begin{minipage}{1.0\linewidth}
        \centering
        \subfloat[RE]{\includegraphics[width=6.5cm,clip]{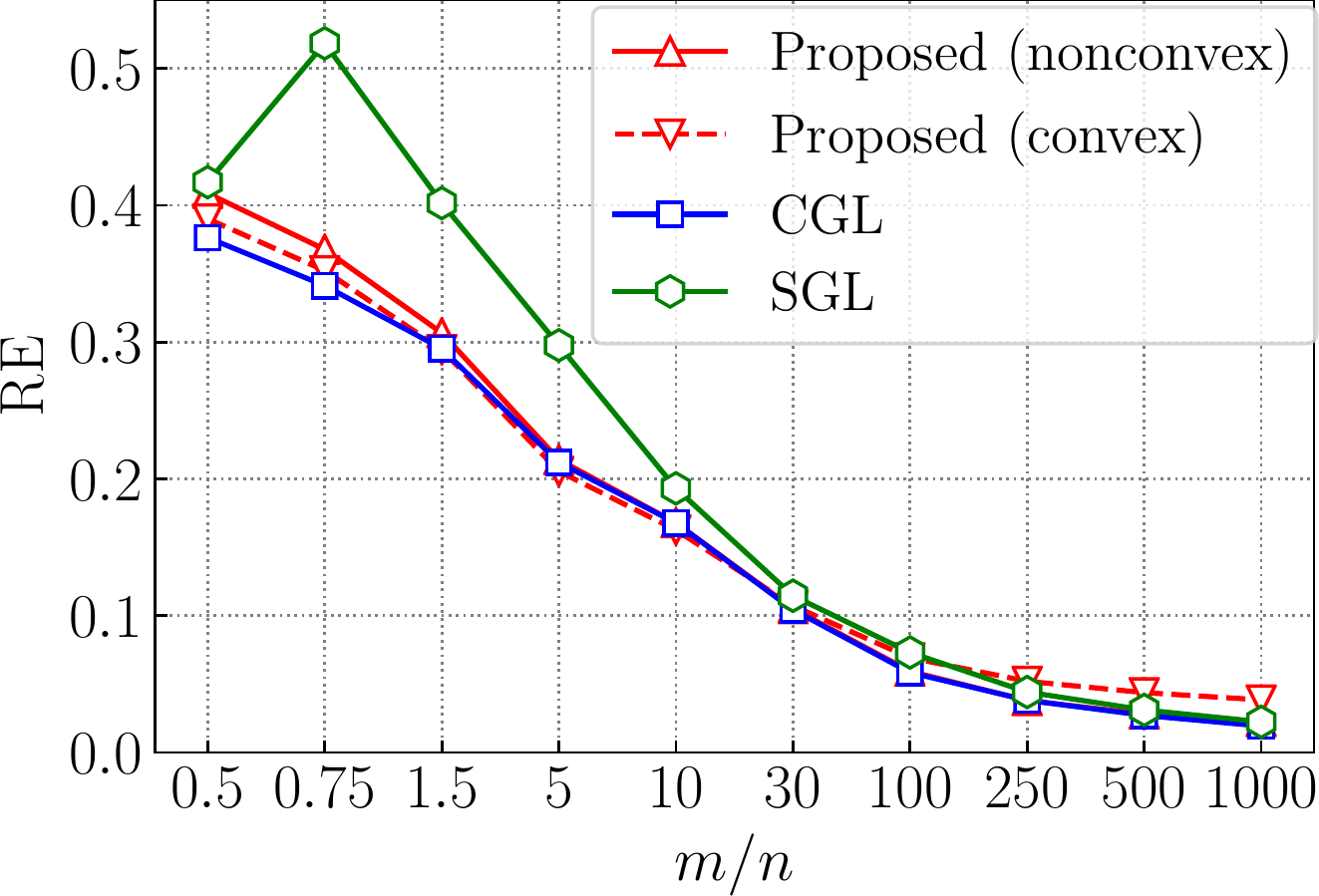}}
    \end{minipage}
    
    \begin{minipage}{1.0\linewidth}
        \centering
        
        \subfloat[FS]{\includegraphics[width=6.5cm,clip]{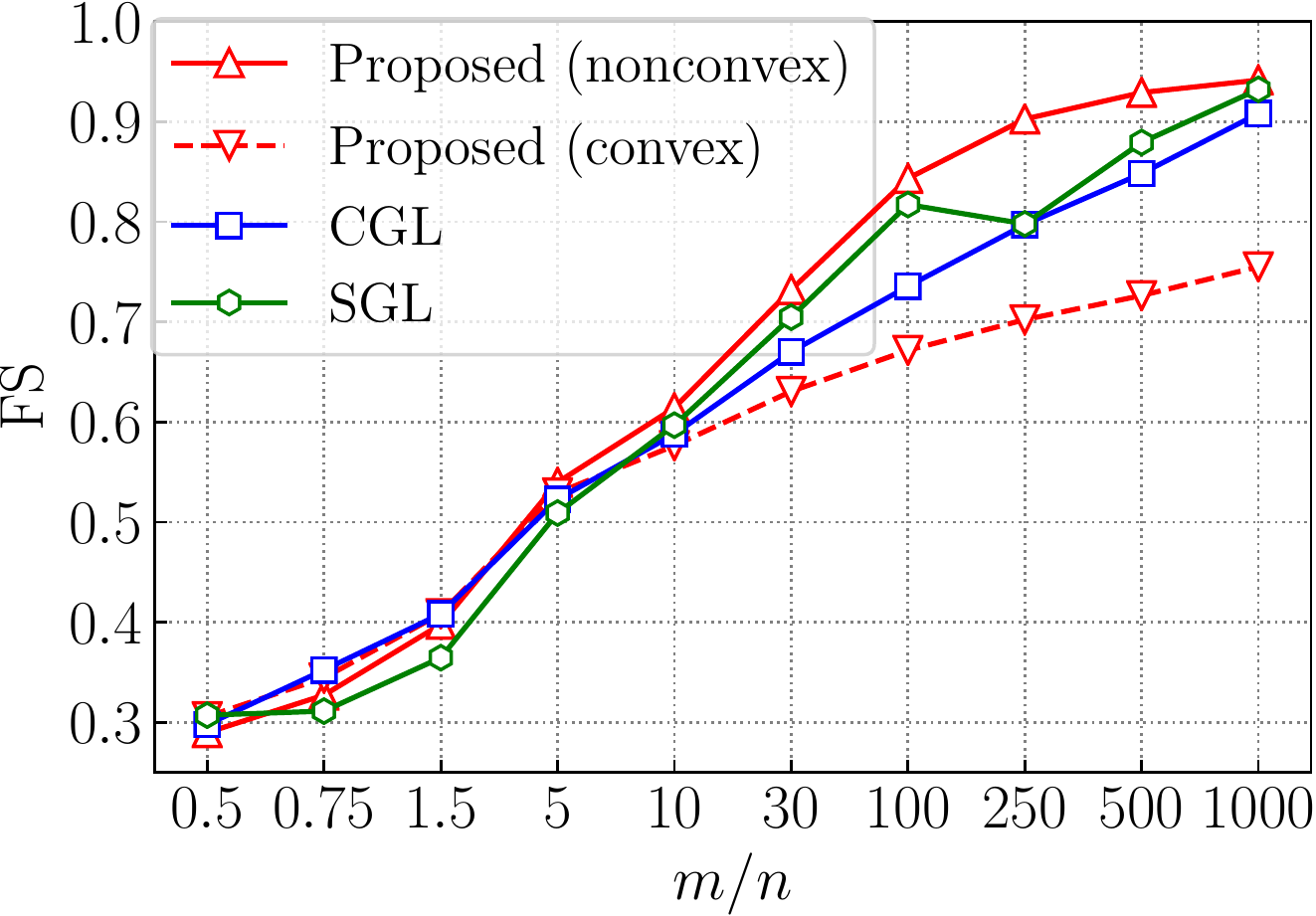}}
    \end{minipage}
        \caption{Experimental results for random modular graph $\mathcal{G}_{\mathrm{M}.}^{(100, 0.01, 0.3)}$}
        \label{modular_RE_FS}
    \end{figure}

\begin{figure}[t]

    \begin{minipage}{1.0\linewidth}
        \centering
        \subfloat[RE]{\includegraphics[width=6.5cm,clip]{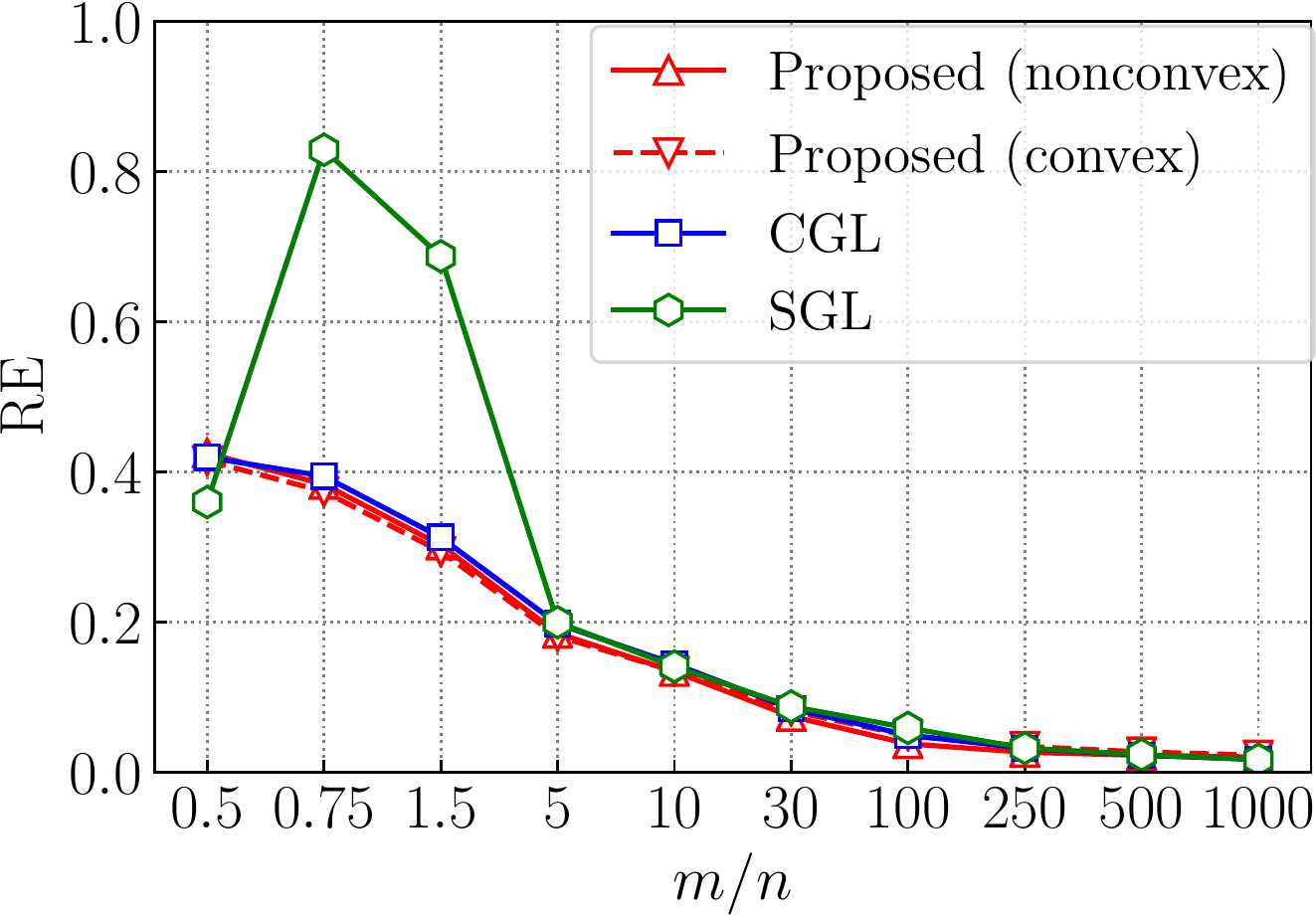}}
    \end{minipage}
    
    \begin{minipage}{1.0\linewidth}
        \centering
        \subfloat[FS]{\includegraphics[width=6.5cm,clip]{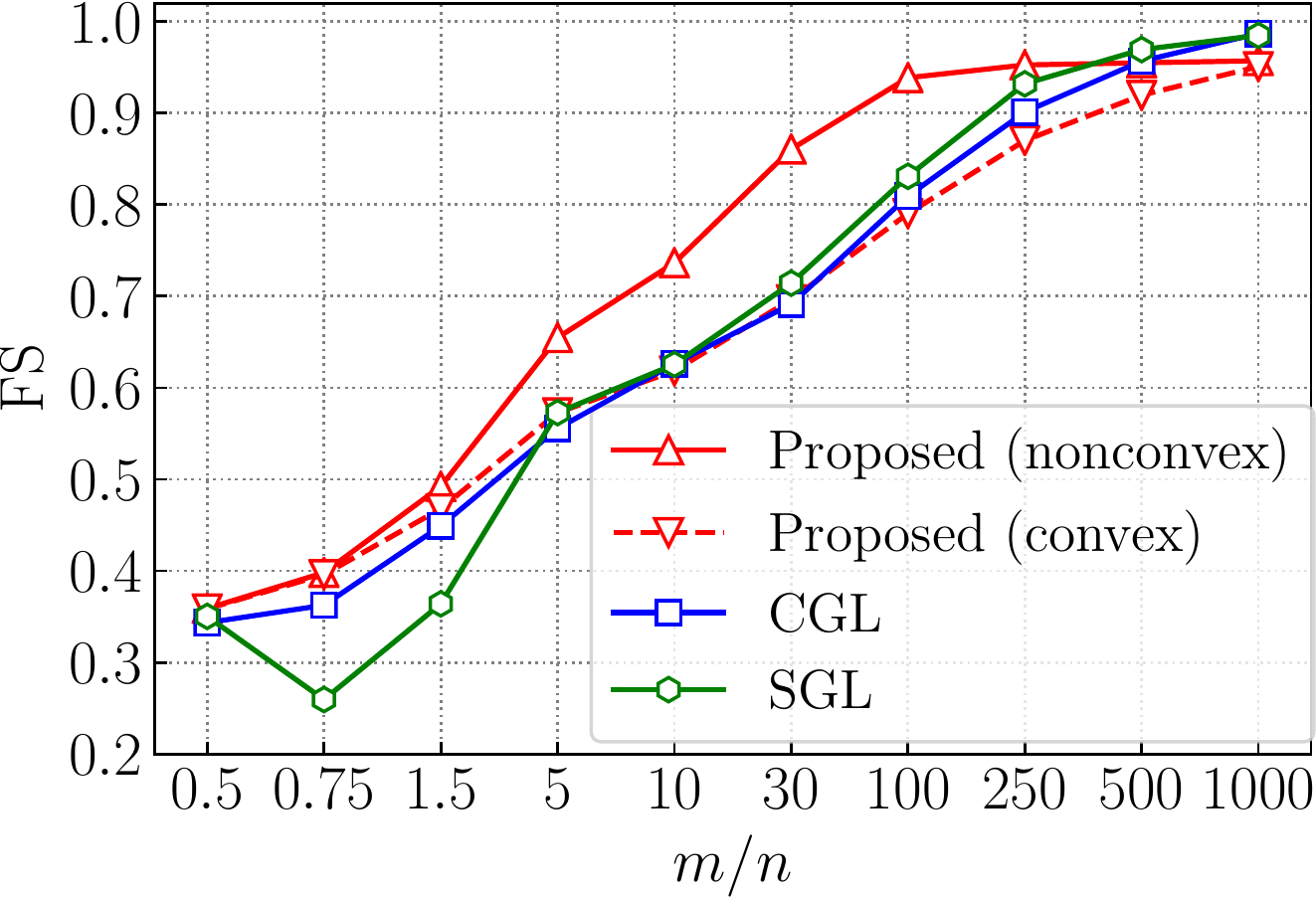}}
    \end{minipage}
        \caption{Experimental results for Erd\"os-R\'{e}nyi graph $\mathcal{G}_{\mathrm{ER}.}^{(100, 0.1)}$}
        \label{ER_RE_FS}
\end{figure}

\noindent{\bf Results:} 
Figure \ref{graph_visualize} shows the ground truth and the learned graph for $m/n=100$ where 
$m$ is the number of measurements.
One can see that the proposed method yields
a more accurate graph than the other methods;
in particular, the graph obtained by the proposed method is remarkably sparse.
Figures \ref{grid_RE_FS} -- \ref{ER_RE_FS} show the performances in RE and FS across $m/n$ for the grid graph $\mathcal{G}_{\mathrm{grid}}^{(10, 10)}$,
the random modular graph $\mathcal{G}_{\mathrm{M}}^{(100, 0.01, 0.3)}$, 
and the Erd\"os-R\'{e}nyi graph $\mathcal{G}_{\mathrm{ER}}^{(100, 0.1)}$,
respectively.
In Fig.~\ref{grid_RE_FS}, the proposed method (nonconvex) significantly outperforms CGL due to the use of the MC penalty, while
the proposed method under the convexity condition achieves approximately the same RE performance as the nonconvex case with the degraded FS performance for large $m/n$.
The performance of SGL is close to that of the proposed method in this case. 
The proposed method under the convexity condition exhibits approximately the same performance as the proposed method for $\lambda_2=0$ (the nonconvex case).
In Fig.~\ref{modular_RE_FS}, on the other hand, the proposed
algorithm outperforms SGL considerably.
The proposed method significantly outperforms CGL in FS 
for large $m/n$, although those two methods exhibit comparable performances in RE (and in FS as well for small $m/n$).
In Fig.~\ref{ER_RE_FS},
the proposed method outperforms the other methods in FS for a wide range of $m/n$ values,
while the proposed method under the convexity condition achieves approximately the same RE performance as the nonconvex case with its FS performance close to those of CGL and SGL for large $m/n$.
We remark that the difference between the proposed method and CGL is more notable in Fig.~\ref{grid_RE_FS} than in Figs.~\ref{modular_RE_FS} and \ref{ER_RE_FS} because the graph in Fig.~\ref{grid_RE_FS} is approximately four times sparser than that of Fig.~\ref{modular_RE_FS} and twice sparser than that of Fig.~\ref{ER_RE_FS}.
Thus, 
the sparsity assumption is a better match to the actual data in for the graph from  Fig.~\ref{grid_RE_FS}. 
We finally remark that, for CGL and SGL, a small regularization parameter was used because use of large regularization parameters with the $\ell_1$ norm leads  to increased errors, which degrade the quality (e.g., interpretability) of the learned graphs.
This is the reason why the graphs obtained by CGL and SGL  in Fig.~\ref{graph_visualize} are not sufficiently sparse.

\subsubsection{Comparisons in computation time}
We investigate how the computation time with tolerance error $1.0\times 10^{-4}$
changes with the size of the graph
for the modular graph $\mathcal{G}_{\mathrm{modular}}^{(n, 0.01, 0.3)}.$
We set $m/n=5000$ and perform graph learning 15 times as in Section \ref{Comparisons_estimate_accuracy}. 
The computation time for $n=160,240,320,400$ is summarized in
Table \ref{ex_graph_size_increase}.
It can be seen that the proposed method is 5.38 -- 43.4 times faster than SGL. Although the computation time of the proposed method is higher due mainly to the eigenvalue decomposition of the larger sized  matrix
compared to CGL, we emphasize that
the performance improvements are remarkable especially for the grid graph (See Fig.~\ref{grid_RE_FS}).
The significant advantage in CPU time is due to the fact that the proposed method requires a few thousand iterations for approximate convergence, while SGL requires over $10^5$ iterations on average
with per-iteration complexity of order $\mathcal{O}(n^3)$.

\begin{table}[t!]
    \centering
    \caption{A comparison of average CPU time (in seconds) for a modular graph $\mathcal{G}_{\mathrm{modular}}^{(n, 0.01, 0.3)}$ with $m/n=5000$.}
\begingroup
\renewcommand{\arraystretch}{1.2}
\begin{tabular}[t]{|c|c|c|c|c|}
\hline
     $n$& proposed method & CGL  & SGL \\\hline 
    160 & 111.2 & 14.95  & 6958 \\\hline
    240 & 427.8 & 51.99  & 40565 \\\hline
    320 & 1601 & 152.6  & 69555 \\\hline
    400 & 4484 & 381.3  & 145613  \\\hline
    \end{tabular}
    \endgroup
    \label{ex_graph_size_increase}
\end{table}
\vspace{-.5em}
\subsection{Experiments with real data}
\begin{figure}[h]
    \begin{center}
        \subfloat[Proposed]{\includegraphics[width=7.5cm,clip]{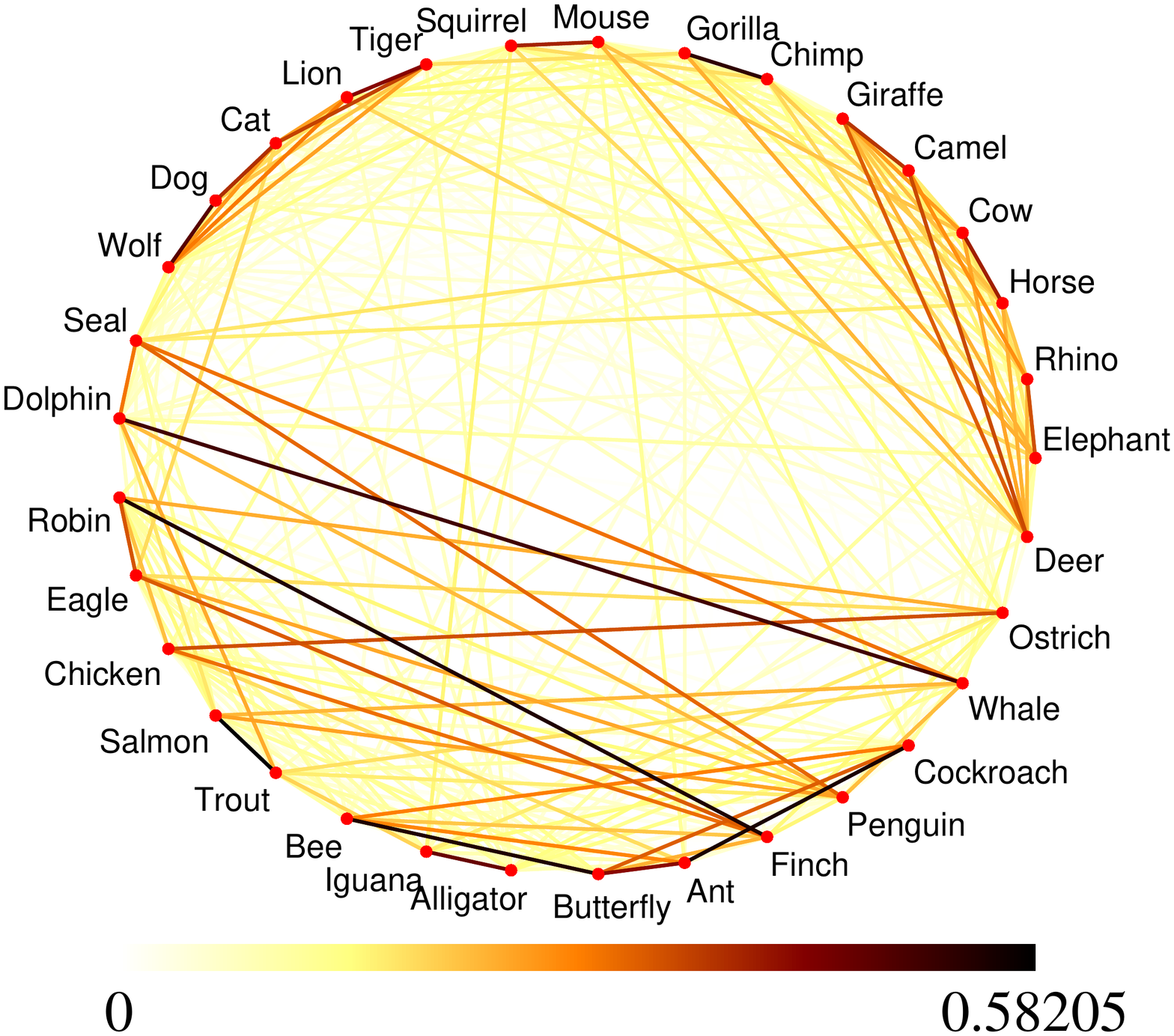}}\\
        \vspace{-1em}
            \subfloat[CGL]{\includegraphics[width=7.5cm,clip]{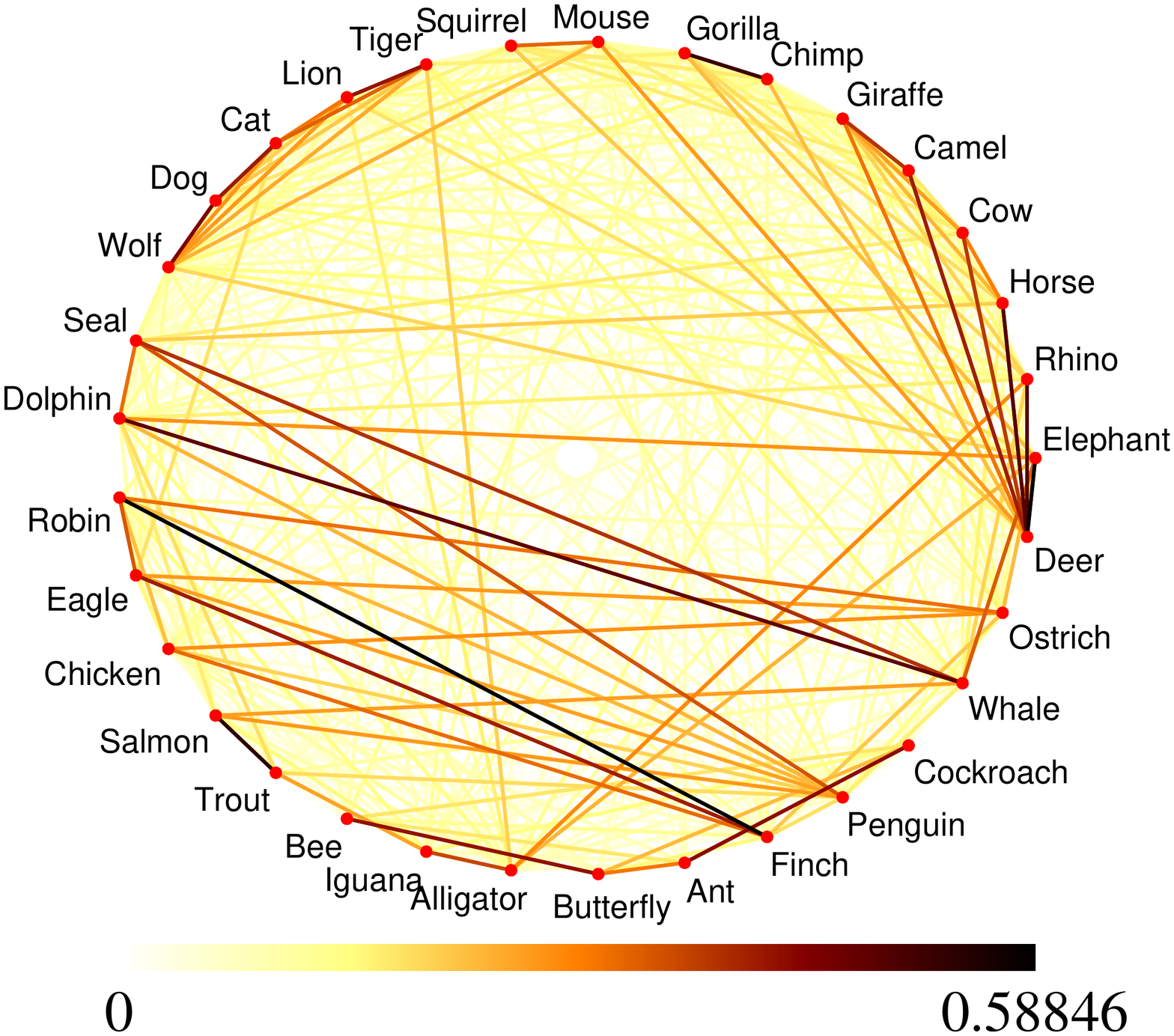}}
            \\
            \vspace{-1em}
            \subfloat[SGL]{\includegraphics[width=7.5cm,clip]{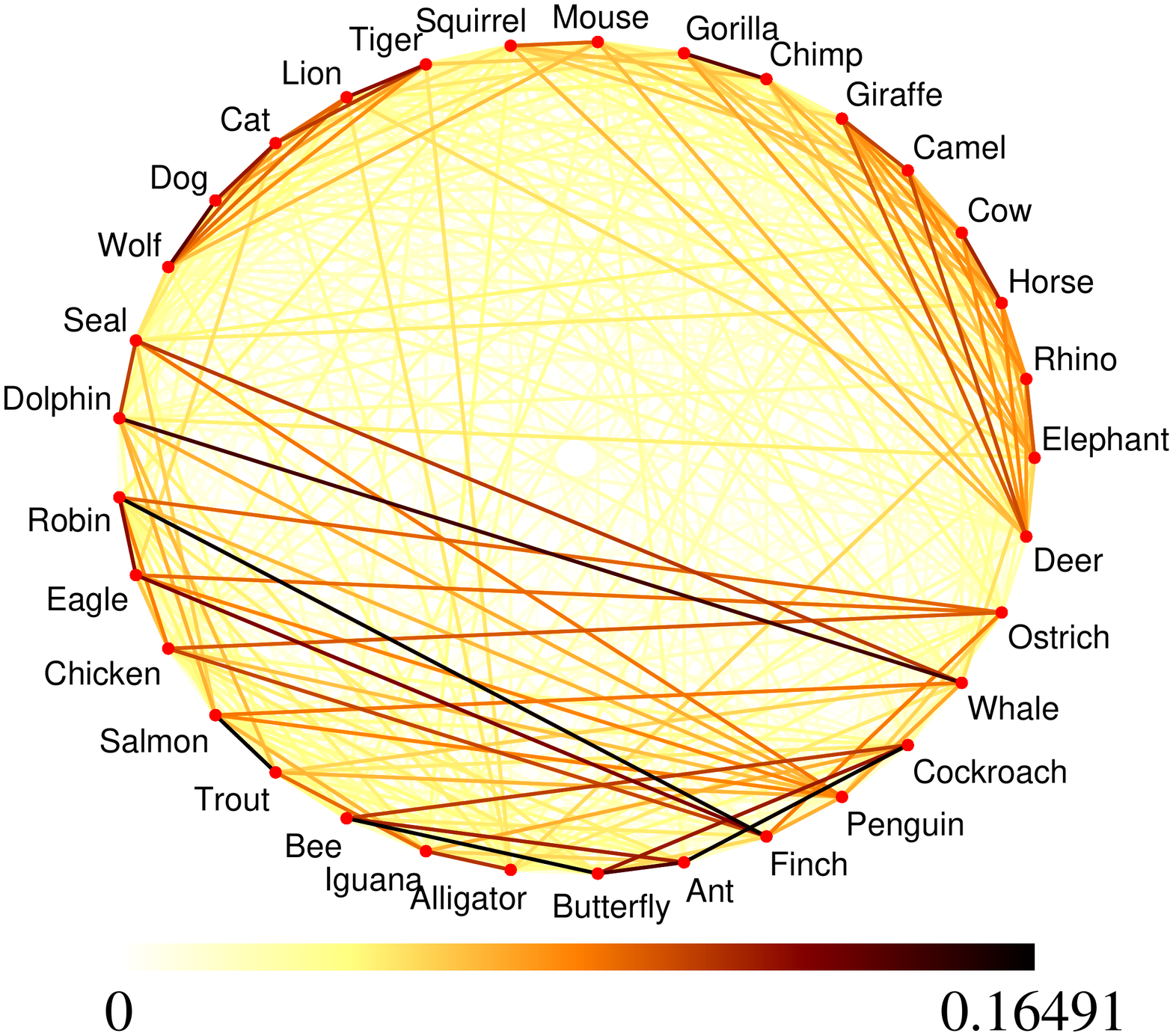}}
    \end{center}
    \caption{The relations of animals from animal dataset.}
     \label{animal_data}
    \end{figure}

We test our method for the animal dataset \cite{Kemp10687},
in which each node represents each animal and the edges represent how much the animals are related to each other.
The dataset contains binary values (i.e., it is a categorical dataset) 
which represent the answers to some questions such as ``has lungs?'' for instance.
There are 102 such questions in total, answered for 33 different animals.
The covariance matrix is created based on this data set, and the graph is learned in the same way as in the previous experiment.
The results are shown in Fig.~\ref{animal_data}.
It can be seen that the proposed method produces a sparser graph than CGL and SGL while preserving the dominant links.
    
\section{Conclusion}
We presented  a graph learning method inserting the nonconvex  MC penalty into the extension of the graphical lasso formulation to produce sparse and accurate graphs.
With the linear-operator-based representation of CGL together with the reformulation 
through the Moreau decomposition, 
an efficient algorithm was derived with
the primal-dual splitting method, for which
an admissible choice of parameters was presented to ensure the provable convergence.
Numerical examples showed that the proposed method significantly outperformed CGL 
for the high-sparsity grid graphs, 
while  still achieving higher F-scores 
than CGL and SGL (the state-of-the-art method)
for the low-sparsity modular graphs. 
In addition, the comparisons in CPU time showed that 
the proposed method was dramatically faster than SGL.

\appendices


\section{Proof of Proposition \ref{pds_def_adjoint}}
\label{appendix_1}
Let $\bm{w}:=[w_{1,2},w_{1,3}, \!\cdots\!, w_{1,n},w_{2,3},\cdots,w_{2,n},\!\cdots\!,w_{n-1,n}]^T$
$\in C$.
By definition of adjoint operator, it holds that
$\langle L(\bm{w}), \bm{M} \rangle$=$\langle L^*(\bm{M}), \bm{w}\rangle$,
of which the left side can be expanded as $w_{1,2}(m_{1,1}+m_{2,2}-m_{1,2}-m_{2,1})+w_{1,3}(m_{1,1}+m_{3,3}-m_{1,3}-m_{3,1})+\cdots$.
It can therefore be seen that 
 $[L^*(\bm{M})]_1= m_{1,1}+m_{2,2}-m_{1,2}-m_{2,1}$ and
  $[L^*(\bm{M})]_2= m_{1,1}+m_{3,3}-m_{1,3}-m_{3,1}$.
In general, it can be verified that $[L^*(\bm{M})]_{(2n-p-1)p/2+q-n}=m_{p,p}+m_{q,q}-m_{p,q}-m_{q,p}$.

\section{Proof of Proposition \ref{prox_inner_prod}}
\label{appendix_2}
\renewcommand{\theequation}{B.\arabic{equation} }
\setcounter{equation}{0}
The assertion can be verified by combining
the basic property \cite[Proposition 24.8]{combettes_book}
\begin{equation}
    \prox_{\phi+\langle \cdot, \bm{u}\rangle}(\bm{x})=\prox_{\phi}(\bm{x}-\bm{u})   \label{prop_1}
\end{equation}
and the following fact \cite[Proposition 1]{yukawa}:
\begin{equation}
    \operatorname{prox}_{\iota_C +\eta\|\cdot\|_1}(\boldsymbol{w})=P_{C}(\boldsymbol{w}-\eta \bm{1}) \label{prop_2}
\end{equation}
for $\eta>0$.
Indeed, since
$ G:=\iota_C +\lambda_1\|\cdot\|_1+\langle \bm{S}, L(\cdot)\rangle=
\iota_C +\lambda_1\|\cdot\|_1+\langle \cdot, L^*(\bm{S})\rangle$,
we have
$\tau G= \iota_C +\tau\lambda_1\|\cdot\|_1+\langle \cdot, \tau L^*(\bm{S})\rangle$.
Applying (\ref{prop_1}\hspace{-0.3em})
to $\phi:=\iota_C +\tau\lambda_1\|\cdot\|_1$ and 
$u=\tau L^*(\bm{S})$, we obtain
\begin{eqnarray}
    \prox_{\tau G}(\bm{w})&=&\prox_{\iota_C +\tau\lambda_1\|\cdot\|_1+\langle \cdot, \tau L^*(\bm{S})\rangle}(\bm{w})\nonumber\\
    &=&\prox_{\iota_C +\tau\lambda_1\|\cdot\|_1}(\bm{w} -  \tau L^*(\bm{S}))
    \nonumber\\
     &=& P_C(\bm{w} -  \tau L^*(\bm{S}) - \tau \lambda_1 \bm{1}).
\end{eqnarray}

\section{Proof of Proposition \ref{prop_prox_logdet}}
\renewcommand{\theequation}{C.\arabic{equation} }
\setcounter{equation}{0}
\label{appendix_3}
The proximity operator of
$$f(\bm{W})\!\mapsto \!\left\{\begin{array}{ll}
    {\!\! -\log \!\operatorname{det} (\bm{W})}, \hspace*{-.8em} &\! {\text { if } \bm{W} \succ 0, } \\
    \!\!{+\infty, } \hspace*{-.8em} & \!{\text { otherwise,}}
    \end{array}\right.$$
    is given by \cite[Example 24.66]{combettes_book}
    \begin{align}
       \begin{aligned}
           &\operatorname{prox}_{-\sigma^{-1}\logdet(\cdot)}(\bm{W})
        \nonumber \\
        &=\frac{1}{2}\bm{Q} \!\operatorname{diag}\!\left(\mu_1\!+\!\sqrt{\mu_{1}^{2}\!+\!4\sigma^{-1}}, \! \cdots \!, \mu_n \!+\!\sqrt{\mu_{n}^{2}\!+\!4\sigma^{-1}}\right) \!\bm{Q}^T.
       \end{aligned}
       \label{C_1}\\
    \end{align}
    By (\ref{C_1}\hspace{-0.25em}) and the property \cite[Proposition 24.8]{combettes_book}:
     \begin{equation}
     \prox_{f(\cdot+\bm{J})}(\bm{W})=\prox_f(\bm{W}+\bm{J})-\bm{J}, \label{C_2}
     \end{equation}
    we obtain the result.
  
\section{Proof of Proposition \ref{convergence_condition}}
\label{appendix_4}
\renewcommand{\theequation}{D.\arabic{equation} }
\setcounter{equation}{0}
It is clear from \eqref{eq:Fw} that $F$ is convex when $\lambda_2\geq \gamma^{-1} \lambda_1$.
The convergence condition of the primal dual splitting method is given as follows \cite{primal-dual}:
\begin{enumerate}
    \item $\frac{1}{\tau} \geq \sigma\|L\|^{2}+\frac{\beta}{2}$, \\
    \item $0<\rho_k<2-\frac{\beta}{2}\left(\frac{1}{\tau}-\sigma\|L\|^{2}\right)^{-1}$,
\end{enumerate}
where $\beta$ is the Lipschitz constant of $\nabla F$ and
$\|L\|:=\sup_{\bm{w}\neq \bm{0}}\frac{\|L(\bm{w})\|_{\rm F}}{\|\bm{w}\|_2}$ is the operator norm.
We show below the Lipschitz constant of $F$ and the operator norm of $L$.
(Although it is shown  in \cite{palomar20} that $\|L\|=\sqrt{2n}$, we show the proof for self-containedness.)

\subsection{Derivation of the Lipschitz constant of $\nabla$ $F$}
The gradient of $F$ is given by
\begin{equation}
    \nabla F(\bm{w})=\gamma^{-1}\lambda_1\prox_{\|\cdot\|_1}(\bm{w})
    + (\lambda_2 - \gamma^{-1} \lambda_1) \bm{w}\nonumber.
\end{equation}
Hence, by the nonexpansiity of the proximity operator as well as the triangular inequality of norm, we obtain
\begin{align}
&\|\nabla F(\bm{w})-\nabla F(\bm{w}')\|_2 \nonumber \\
&=\left\|
\left(
\gamma^{-1}\lambda_1\prox_{\|\cdot\|_1}(\bm{w})
    \!+\! (\lambda_2 - \gamma^{-1} \lambda_1) \bm{w}
\right)
\right. \nonumber\\
&\quad \left.-\left(
\gamma^{-1}\lambda_1\prox_{\|\cdot\|_1}(\bm{w}')
    \!+\! (\lambda_2 - \gamma^{-1} \lambda_1) \bm{w}'
\right) \right\|_2\\
&\leq \|\gamma^{-1}\lambda_1\prox_{\|\cdot\|_1}(\bm{w}) - \gamma^{-1}\lambda_1\prox_{\|\cdot\|_1}(\bm{w}')\|_2 \nonumber\\
&\quad+ 
\|
(\lambda_2 - \gamma^{-1} \lambda_1) (\bm{w} - \bm{w}')
\|_2 \\
&\leq \gamma^{-1}\lambda_1\|\bm{w}-\bm{w}'\|_2 + (\lambda_2 - \gamma^{-1}\lambda_1)\|\bm{w}-\bm{w}'\|_2 \\
&\leq \lambda_2 \|\bm{w}-\bm{w}'\|_2,
\end{align}
from which $\nabla F$ is $\lambda_2$-Lipschitz contiuous.

 \subsection{Derivation of $\|L\|$}
By definition of Laplacian, we have $[L(\bm{w})]_{i,i} = - \sum_{j\neq i} [L(\bm{w})]_{i,j}$,
and hence it holds that
\begin{align}
&\|L\|^2=\sup_{\bm{w}\neq \bm{0}}\frac{\|L(\bm{w})\|_{\rm F}^2}{\|\bm{w}\|_2^2} \nonumber \\
&=\sup_{\bm{w}\neq \bm{0}}\frac{\displaystyle\sum_{p=1}^n \left[ \left(-\displaystyle\sum_{q\neq p} [L(\bm{w})]_{p, q}\right)^2\right]+2\left(\sum_{p < q} [L(\bm{w})]_{p, q}^2\right) }{\displaystyle\sum_{p < q} [L(\bm{w})]_{p, q}^2} \nonumber\\
&=\sup_{\bm{w}\neq \bm{0}}\frac{\displaystyle\sum_{p=1}^n\left[\left(-\sum_{q\neq p} [L(\bm{w})]_{p, q}\right)^2\right]}{\displaystyle\sum_{p < q} [L(\bm{w})]_{p, q}^2}+2. \label{product_sum}
\end{align}
Using the Caucy-Schwartz inequality, we have
\begin{align}
    \sum_{p=1}^n\left[\left(-\sum_{q\neq p} [L(\bm{w})]_{p,q}  \right)^2\right] \!&\!\leq\! \sum_{p=1}^n \left[(n-1)\sum_{q \neq p} [L(\bm{w})]_{p, q}^2\right] \nonumber \\
    &=2(n-1)\!\sum_{p < q}  [L(\bm{w})]_{p, q}^2,
    \label{product_sum_2}
\end{align}
where the inequality holds with equality when $\bm{w}=\alpha\bm{1}$, $\alpha\in\mathbb{R}$.
By (\ref{product_sum}\hspace{-0.25em}) and (\ref{product_sum_2}\hspace{-0.25em}), we obtain
$\|L\|^2=2n$, where
the upper bound is obtained when we consider the complete graph with all weights equal to one; i.e., $\bm{w}=\bm{1}$.


\printbibliography

@article{ortega16,
  author   = {H. E. {Egilmez} and E. {Pavez} and A. {Ortega}},
  journal  = {IEEE Journal of Selected Topics in Signal Processing},
  title    = {Graph Learning From Data Under {Laplacian} and Structural Constraints},
  year     = {2017},
  volume   = {11},
  number   = {6},
  pages    = {825-841},
  doi      = {10.1109/JSTSP.2017.2726975},
  issn     = {1941-0484},
  month   = {Sep.}
}

@book{combettes_book,
  author  = {Heinz H. {Bauschke} and Patrick L. {Combettes}},
  Publication = {Springer},
  title   = {Convex Analysis and Monotone Operator Theory in Hilbert Spaces},
  year    = {2017},
  edition = {Second}
}

@article{primal-dual,
  author  = {Condat, L},
  journal = {Journal of Optimization Theory and Applications},
  title   = {A Primal–Dual Splitting Method for Convex Optimization Involving {Lipschitzian}, Proximable and Linear Composite Terms},
  year    = {2013},
  volume  = {158},
  pages   = {460-479},
  doi     = {10.1007/s10957-012-0245-9}
}

@article{yukawa,
  title     = {Supervised nonnegative matrix factorization via minimization of regularized {{Moreau}}-envelope of divergence function with application to music transcription},
  author    = {Masahiro Yukawa and Hideaki Kagami},
  year      = {2018},
  month    = {Mar.},
  day       = {1},
  doi       = {10.1016/j.jfranklin.2017.12.002},
  volume    = {355},
  pages     = {2041--2066},
  journal   = {Journal of the Franklin Institute},
  issn      = {0016-0032},
  publisher = {Elsevier Limited},
  number    = {4}
}

@article{MagoarouGT16,
  author  = {L. {Le Magoarou} and R. {Gribonval} and N. {Tremblay}},
  journal = {IEEE Transactions on Signal and Information Processing over Networks},
  title   = {Approximate Fast Graph {fourier} Transforms via Multilayer Sparse Approximations},
  year    = {2018},
  volume  = {4},
  number  = {2},
  pages   = {407-420},
  doi     = {10.1109/TSIPN.2017.2710619}
}

@article{connect,
  author   = {G. {Mateos} and S. {Segarra} and A. G. {Marques} and A. {Ribeiro}},
  journal  = {IEEE Signal Processing Magazine},
  title    = {Connecting the Dots: Identifying Network Structure via Graph Signal Processing},
  year     = {2019},
  volume   = {36},
  number   = {3},
  pages    = {16-43},
  doi      = {10.1109/MSP.2018.2890143},
  issn     = {1558-0792},
  month   = {May.}
}

@article{selesnick,
  author   = {I. {Selesnick}},
  journal  = {IEEE Transactions on Signal Processing},
  title    = {Sparse Regularization via Convex Analysis},
  year     = {2017},
  volume   = {65},
  number   = {17},
  pages    = {4481-4494},
  doi      = {10.1109/TSP.2017.2711501},
  issn     = {1941-0476},
  month   = {Sep.}
}

@article{friedman2008sparse,
  title     = {Sparse inverse covariance estimation with the graphical lasso},
  author    = {Friedman, Jerome and Hastie, Trevor and Tibshirani, Robert},
  journal   = {Biostatistics},
  volume    = {9},
  number    = {3},
  pages     = {432--441},
  year      = {2008},
  publisher = {Oxford University Press}
}

@article{Lake,
  author  = {Lake, Brenden and Tenenbaum, Joshua},
  journal = {Proceedings of 33rd Annual Conference of the Cognitive Science Society},
  title   = {Discovering Structure by Learning Sparse Graphs},
  year    = {2010},
  pages   = {778-783},
  month  = {Aug.}
}

@inproceedings{HallacPBL17,
  author    = {Hallac, David and Park, Youngsuk and Boyd, Stephen and Leskovec, Jure},
  title     = {Network Inference via the Time-Varying Graphical Lasso},
  year      = {2017},
  isbn      = {9781450348874},
  doi       = {10.1145/3097983.3098037},
  booktitle = {Proceedings of the 23rd ACM SIGKDD International Conference on Knowledge Discovery and Data Mining},
  pages     = {205--213},
  numpages  = {9},
}

@article{palomar20,
  author  = {S. {Kumar} and J. {Ying} and J. V. {de Miranda Cardoso}  and D. P. {Palomar}},
  year    = {2020},
  title   = {A unified framework for structured graph learning via spectral constraints},
  journal = {Journal of Machine Learning Research},
  volume  = {21},
  pages   = {1--60}
}

@article{Shen12,
  author  = {X. {Shen} and  W. {Pan} and Y. {Zhu}},
  title   = {Likelihood-based selection and sharp parameter estimation},
  journal = {Journal of the American Statistical Association},
  year    = {2012},
  month      = {Jan.},
  number = {497},
  volume  = {107},
  pages   = {223--232},
  doi     = {10.1080/01621459.2011.645783}
}

@article{lam2009,
  author    = {Lam, Clifford and Fan, Jianqing},
  doi       = {10.1214/09-AOS720},
  fjournal  = {Annals of Statistics},
  journal   = {Annals of Statistics},
  month    = {Dec.},
  number    = {6B},
  pages     = {4254--4278},
  publisher = {The Institute of Mathematical Statistics},
  title     = {Sparsistency and rates of convergence in large covariance matrix estimation},
  url       = {https://doi.org/10.1214/09-AOS720},
  volume    = {37},
  year      = {2009}
}

@article{VECCHIO2017206,
  title    = {Connectome: Graph theory application in functional brain network architecture},
  journal  = {Clinical Neurophysiology Practice},
  volume   = {2},
  pages    = {206 - 213},
  year     = {2017},
  issn     = {2467-981X},
  doi      = {https://doi.org/10.1016/j.cnp.2017.09.003},
  url      = {http://www.sciencedirect.com/science/article/pii/S2467981X17300276},
  author   = {Fabrizio Vecchio and Francesca Miraglia and Paolo Maria {Rossini}}
}

@article{Mason2007,
  author    = {O. Mason and  M. Verwoerd},
  issn      = {1751-8849},
  title     = {Graph theory and networks in Biology},
  journal   = {IET Systems Biology},
  volume    = {1},
  year      = {2007},
  month    = {Mar.},
  pages     = {89-119},
  number= {30},
  publisher = {Institution of Engineering and Technology},
  copyright = {© The Institution of Engineering and Technology},
  url       = {https://digital-library.theiet.org/content/journals/10.1049/iet-syb_20060038}
}

@article{Bhuyan16,
  author  = {M. H. {Bhuyan} and D. K. {Bhattacharyya} and J. K. {Kalita}},
  journal = {IEEE Communications Surveys Tutorials},
  title   = {Network Anomaly Detection Methods, Systems and Tools},
  year    = {2014},
  volume  = {16},
  number  = {1},
  pages   = {303-336}
}

@article{Bhuyan14,
  author  = {P. {Danaher} and P. {Wang} and D. M. {Witten}},
  journal = {Journal of Royal Statistical Society. Series B: Statistical Methodological},
  title   = {The joint graphical lasso for inverse covariance estimation across multiple classes},
  year    = {2014},
  volume  = {76},
  number  = {2},
  pages   = {373-397}
}

@article{wang2012,
  author    = {Wang, Hao},
  doi       = {10.1214/12-BA729},
  fjournal  = {Bayesian Analysis},
  journal   = {Bayesian Analysis},
  month    = {Dec.},
  number    = {4},
  pages     = {867--886},
  publisher = {International Society for {Bayesian} Analysis},
  title     = {{Bayesian} Graphical Lasso Models and Efficient Posterior Computation},
  url       = {https://doi.org/10.1214/12-BA729},
  volume    = {7},
  year      = {2012}
}

@article{Sun12,
  author  = {Shiliang Sun  and Rongqing Huang  and Ya Gao },
  title   = {Network-Scale Traffic Modeling and Forecasting with Graphical Lasso and Neural Networks},
  journal = {Journal of Transportation Engineering},
  volume  = {138},
  number  = {11},
  pages   = {1358-1367},
  year    = {2012},
  doi     = {10.1061/(ASCE)TE.1943-5436.0000435},
  url     = {https://ascelibrary.org/doi/abs/10.1061/%28ASCE%29TE.1943-5436.0000435},
  eprint  = {https://ascelibrary.org/doi/pdf/10.1061/%28ASCE%29TE.1943-5436.0000435}
}

@article{zhang2010,
  author    = {Zhang, Cun Hui},
  doi       = {10.1214/09-AOS729},
  fjournal  = {Annals of Statistics},
  journal   = {Annals of Statistics},
  month    = {Apr.},
  number    = {2},
  pages     = {894--942},
  publisher = {The Institute of Mathematical Statistics},
  title     = {Nearly unbiased variable selection under minimax concave penalty},
  url       = {https://doi.org/10.1214/09-AOS729},
  volume    = {38},
  year      = {2010}
}

@book{GMRF,
  author  = {H. Rue and L. Held},
  journal = {Chapman \& Hall/CRC},
  title   = {{Gaussian} {Markov} random fields: theory and applications},
  edition = {First},
  year    = {2005}
}

@article{mazumder2012,
  author    = {Mazumder, Rahul and Hastie, Trevor},
  doi       = {10.1214/12-EJS740},
  fjournal  = {Electronic Journal of Statistics},
  journal   = {Electronic Journal of Statistics},
  pages     = {2125--2149},
  publisher = {The Institute of Mathematical Statistics and the Bernoulli Society},
  title     = {The graphical lasso: New insights and alternatives},
  url       = {https://doi.org/10.1214/12-EJS740},
  volume    = {6},
  year      = {2012}
}

@article{JMLR:v9:banerjee08a,
  author  = {Onureena Banerjee and Laurent El Ghaoui and Alexandre d'Aspremont},
  title   = {Model Selection Through Sparse Maximum Likelihood Estimation for Multivariate {Gaussian} or Binary Data},
  journal = {Journal of Machine Learning Research},
  year    = {2008},
  volume  = {9},
  number  = {15},
  pages   = {485-516},
  url     = {http://jmlr.org/papers/v9/banerjee08a.html}
}

@article{perraudin2014gspbox,
  author        = {{Perraudin}, Nathana{\"e}l and {Paratte}, Johan and {Shuman}, David and {Martin}, Lionel  and {Kalofolias}, Vassilis and 
	{Vandergheynst}, Pierre and {Hammond}, David K. },
  title         = {{GSPBOX: A toolbox for signal processing on graphs}},
  journal       = {ArXiv e-prints},
  archiveprefix = {arXiv},
  eprint        = {1408.5781},
  primaryclass  = {cs.IT},
  year          = 2014,
  month         = {Aug.},
  adsurl        = {http://arxiv.org/abs/1408.5781}
}

@article{Abe_2020,
  title     = {Linearly involved generalized {{Moreau}} enhanced models and their proximal splitting algorithm under overall convexity condition},
  volume    = {36},
  issn      = {1361-6420},
  url       = {http://dx.doi.org/10.1088/1361-6420/ab551e},
  doi       = {10.1088/1361-6420/ab551e},
  number    = {3},
  journal   = {Inverse Problems},
  publisher = {IOP Publishing},
  author    = {Abe, Jiro and Yamagishi, Masao and Yamada, Isao},
  year      = {2020},
  month     = {Feb.},
  pages     = {1-36}
}

@article{suzuki21,
  author  = {K.~Suzuki and M.~Yukawa},
  title   = {Robust recovery of jointly-sparse signals using minimax concave loss function},
  journal = {IEEE Transactions on Signal Process.},
  volume  = {69},
  pages   = {669--681},
  year    = {2020},
  month   = {Dec.}
}

@inproceedings{kaneko20,
  author    = {Hiroyuki Kaneko and Masahiro Yukawa},
  title     = {Normalized least-mean-square algorithms with minimax concave penalty},
  booktitle = {Proceedings of 45th International Conference on Acoustics, Speech and Signal Processing (ICASSP)},
  year      = 2020,
  pages     = {5440--5444}
}

@article{yukawa21_alime,
  author = {M.~Yukawa and H.~Kaneko and K.~Suzuki and I.~Yamada},
  title  = {Linearly-involved {Moreau}-Enhanced-over-Subspace Model: Debiased Sparse Modeling and Stable Outlier-Robust Regression},
  year   = 2021,
  note   = {submitted for publication}
}

@inproceedings{koyakumaru_21,
  author    = {Tatsuya Koyakumaru and Masahiro Yukawa and Eduardo Pavez and Antonio Ortega},
  title     = {A Graph Learning Algorithm Based on {Gaussian} {Markov} Random Fields and Minimax Concave Penalty},
  booktitle = {Proceedings of 46th International Conference on Acoustics, Speech and Signal Processing (ICASSP)},
  year      = 2021,
   pages = {5390--5394},
  %note   = {submitted for publication}
}

@Manual{koyakumaru_bthesis,
author = {T.~Koyakumaru},
  title = 	 {Graph learning based on {Gaussian} {Markov} random fields and minimax concave penalty},
  organization = {Keio University},
    OPTmonth = 	 {Feb.},
  year = 	 {2020},
  note = 	 {Bachelor Thesis (in Japanese)},
  }

@article {Kemp10687,
	author = {Kemp, Charles and Tenenbaum, Joshua B.},
	title = {The discovery of structural form},
	volume = {105},
	number = {31},
	pages = {10687--10692},
	year = {2008},
	doi = {10.1073/pnas.0802631105},
	publisher = {National Academy of Sciences},
	issn = {0027-8424},
	URL = {https://www.pnas.org/content/105/31/10687},
	eprint = {https://www.pnas.org/content/105/31/10687.full.pdf},
	journal = {Proceedings of the National Academy of Sciences}
}

@book{Buhlmann_book,
  author  = {P. {Buhlmann} and S. van. de. {Geer}},
  publication = {Springer},
  title   = {Statistics for High-Dimensional Data:Methods, Theory and Applications},
  year    = {2011},
}

@ARTICLE{GSP_GL_2,
  author={S. {Segarra} and A. G. {Marques} and G. {Mateos} and A. {Ribeiro}},
  journal={IEEE Transactions on Signal and Information Processing over Networks}, 
  title={Network Topology Inference from Spectral Templates}, 
  year={2017},
  volume={3},
  number={3},
  pages={467-483},
  doi={10.1109/TSIPN.2017.2731051}}

@ARTICLE{Dong_servey,
  author={X. {Dong} and D. {Thanou} and M. {Rabbat} and P. {Frossard}},
  journal={IEEE Signal Processing Magazine}, 
  title={Learning Graphs From Data: A Signal Representation Perspective}, 
  year={2019},
  volume={36},
  number={3},
  pages={44-63},
  doi={10.1109/MSP.2018.2887284}}

@ARTICLE{Meinshausen_06,
  author={N. {Meinshausen} and P. {Buhlmann}},
  journal={The Annals of Statistics}, 
  title={High-dimensional graphs and variable
selection with the Lasso}, 
  year={2006},
  volume={34},
  pages={1436–1462}}

@INPROCEEDINGS{tanaka19_time_varying,
  author={Yamada, Koki and Tanaka, Yuichi and Ortega, Antonio},
  booktitle={Proceedings of 44th International Conference on Acoustics, Speech and Signal Processing (ICASSP)}, 
  title={Time-varying Graph Learning Based on Sparseness of Temporal Variation}, 
  year={2019},
  volume={},
  number={},
  pages={5411-5415},
  doi={10.1109/ICASSP.2019.8682762}}

@ARTICLE{ex_MCP_1_Feature_Selection,
  author={Laporte, Léa and Flamary, Rémi and Canu, Stéphane and Déjean, Sébastien and Mothe, Josiane},
  journal={IEEE Transactions on Neural Networks and Learning Systems}, 
  title={Nonconvex Regularizations for Feature Selection in Ranking With Sparse SVM}, 
  year={2014},
  volume={25},
  number={6},
  pages={1118-1130},
  doi={10.1109/TNNLS.2013.2286696}}

@ARTICLE{ex_MCP_2_Bearing_Fault_Diagnosis,
  author={Wang, Shibin and Selesnick, Ivan and Cai, Gaigai and Feng, Yining and Sui, Xin and Chen, Xuefeng},
  journal={IEEE Transactions on Industrial Electronics}, 
  title={Nonconvex Sparse Regularization and Convex Optimization for Bearing Fault Diagnosis}, 
  year={2018},
  volume={65},
  number={9},
  pages={7332-7342},
  doi={10.1109/TIE.2018.2793271}}

@article{Nurminskii_73_weakly_convex,
author = {Nurminskii, E.A.} ,
title = {The quasigradient method for the solving of the nonlinear programming problems} ,
journal={Cybernetics and Systems Analysis},
volume={9},
pages={145-–150},
year={1973},
}

@article{Mazumder_11, 
author = {Rahul Mazumder and Jerome H. Friedman and Trevor Hastie},
title = {SparseNet: Coordinate Descent With Nonconvex Penalties},
journal = {Journal of the American Statistical Association},
volume = {106},
number = {495},
pages = {1125-1138},
year  = {2011},
publisher = {Taylor & Francis},
doi = {10.1198/jasa.2011.tm09738},
URL = {https://doi.org/10.1198/jasa.2011.tm09738
},
eprint = {https://doi.org/10.1198/jasa.2011.tm09738}
}

@ARTICLE{surveyOfNonconvexPenalty,
  author={Wen, Fei and Chu, Lei and Liu, Peilin and Qiu, Robert C.},
  journal={IEEE Access}, 
  title={A Survey on Nonconvex Regularization-Based Sparse and Low-Rank Recovery in Signal Processing, Statistics, and Machine Learning}, 
  year={2018},
  volume={6},
  number={},
  pages={69883--69906},
  doi={10.1109/ACCESS.2018.2880454}}

@book{book_Probabilistic_graphical_model,
author = {Koller, Daphne and Friedman, Nir},
title = {Probabilistic Graphical Models: Principles and Techniques - Adaptive Computation and Machine Learning},
year = {2009},
isbn = {0262013193},
publisher = {The MIT Press}
}

@inproceedings{palomar_MC,
 author = {Ying, Jiaxi and de Miranda Cardoso , Jos\'{e} Vin\'{\i}cius and Palomar, Daniel},
 booktitle = {Advances in Neural Information Processing Systems},
 pages = {7101--7113},
 title = {Nonconvex Sparse Graph Learning under Laplacian Constrained Graphical Model},
 url = {https://proceedings.neurips.cc/paper/2020/file/4ef42b32bccc9485b10b8183507e5d82-Paper.pdf},
 volume = {33},
 year = {2020}
}

@article{zhang2020learning,
      title={Learning Graph Laplacian with {MCP}}, 
      author={Yangjing Zhang and Kim Chuan Toh and Defeng Sun},
      year={2020},
      eprint={2010.11559},
      journal = {ArXiv e-prints},
      primaryClass={cs.LG}
}

@inproceedings{komuro_2021_SSP,
  author={Komuro, Kei and Yukawa, Masahiro and Cavalcante, Renato L. G.},
  booktitle={2021 IEEE Statistical Signal Processing Workshop (SSP)}, 
  title={Distributed Sparse Optimization With Minimax Concave Regularization}, 
  year={2021},
  volume={},
  number={},
  pages={31-35}
 }

\vfill
\EOD

\end{document}